\documentclass[aps, pre, reprint, superscriptaddress, onecolumn, tightenlines,
12pt, notitlepage, longbibliography]{revtex4-1}
\usepackage{amsmath}
\usepackage{graphicx}
\usepackage{bm}
\usepackage{microtype}

\renewcommand{\vec}[1]{\bm{#1}}
\newcommand{\uvec}[1]{\bm{\hat{#1}}}
\newcommand{\bra}[1]{\left\langle#1\right|}
\newcommand{\ket}[1]{\left|#1\right\rangle}
\newcommand{\avr}[1]{\left\langle#1\right\rangle}

\newcommand{\Lv}{\mathcal{L}}
\newcommand{\U}{\mathsf{U}}
\renewcommand{\P}{\mathcal{P}}
\newcommand{\Q}{\mathcal{Q}}

\DeclareMathOperator{\Pe}{Pe}
\DeclareMathOperator{\Kn}{Kn}

\begin{document}
\title{Integration through transients for inelastic hard sphere fluids}
\author{W. Till Kranz}
\email{kranz@thp.uni-koeln.de}
\affiliation{Institut f\"ur Theoretische Physik, Universit\"at zu
  K\"oln, 50937 K\"oln, Germany}
\affiliation{Institut f\"ur Materialphysik im Weltraum, Deutsches
  Zentrum f\"ur Luft- und Raumfahrt, 51170 K\"oln, Germany}
\author{Fabian Frahsa}
\affiliation{Fachbereich Physik, Universit\"at Konstanz, 78457
  Konstanz, Germany}
\author{Annette Zippelius}
\affiliation{Institut f\"ur Theoretische Physik,
  Georg-August-Universit\"at G\"ottingen, 37077 G\"ottingen, Germany}
\author{Matthias Fuchs}
\affiliation{Fachbereich Physik, Universit\"at Konstanz, 78457
  Konstanz, Germany}
\author{Matthias Sperl}
\affiliation{Institut f\"ur Materialphysik im Weltraum, Deutsches
  Zentrum f\"ur Luft- und Raumfahrt, 51170 K\"oln, Germany}
\affiliation{Institut f\"ur Theoretische Physik, Universit\"at zu
  K\"oln, 50937 K\"oln, Germany}

\begin{abstract}
  We compute the rheological properties of inelastic hard spheres in
  steady shear flow for general shear rates and densities. Starting
  from the microscopic dynamics we generalise the Integration Through
  Transients (\textsc{itt}) formalism to a fluid of dissipative,
  randomly driven granular particles. The stress relaxation function
  is computed approximately within a mode-coupling theory---based on
  the physical picture, that relaxation of shear is dominated by slow
  structural relaxation, as the glass transition is approached. The
  transient build-up of stress in steady shear is thus traced back to
  transient density correlations which are computed self-consistently
  within mode-coupling theory. The glass transition is signalled by
  the appearance of a yield stress and a divergence of the Newtonian
  viscosity, characterizing linear response. For shear rates
  comparable to the structural relaxation time, the stress becomes
  independent of shear rate and we observe shear thinning, while for
  the largest shear rates Bagnold scaling, i.e., a quadratic increase of
  shear stress with shear rate, is recovered. The rheological
  properties are qualitatively similar for all values of
  $\varepsilon$, the coefficient of restitution; however, the
  magnitude of the stress as well as the range of shear thinning and
  thickening show significant dependence on the inelasticity.
\end{abstract}

\maketitle

\section{Introduction}
\label{sec:introduction}

Understanding the rheology of granular fluids from a kinetic theory
point of view has been of interest from the earliest days of granular
physics
\cite{bagnold54,haff83,walton+braun86,campbell90,jaeger+nagel92,iverson97,brey+dufty98,savage98}. It
has been established that the granular Boltzmann, or Boltzmann-Enskog
equation \cite{brilliantov+poeschel10,goldshtein+shapiro95} is a
useful starting point to understand granular gases at low density
\cite{campbell90,goldshtein+shapiro95,brey+dufty98,goldhirsch03,brilliantov+poeschel10,vollmayr+aspelmeier11}. Various
methods to extract the transport coefficients in linear response,
originally developed for molecular gases have been generalized to
non-equilibrium granular gases
\cite{brey+dufty98,garzo+montanero02,jenkins+richman85,garzo13}. The
general understanding is that the Navier-Stokes equations of
hydrodynamics (with small modifications \cite{dufty07}) provide a
useful description of granular gas flow. Unfortunately, the range of
natural phenomena involving flow at low density and infinitesimal
perturbations is much smaller for granular media than it is for
classical gas flow. Most granular flows occur at both high volume
fraction and significant shear rates. Indeed every gravity driven
granular flow will start from a granular solid at rest
\cite{jaeger+nagel96,iverson97,kokelaar+bahia17} with a packing
fraction around the random close packing density of the respective
material. Only recently, first proposals to extend granular kinetic
theory to relevant densities have appeared
\cite{reddy+kumaran07,kumaran14,suzuki+hayakawa14,kranz+sperl17,jenkins17+berzi}.

\begin{figure}[t]
  \centering
  \includegraphics[width=6.4in]{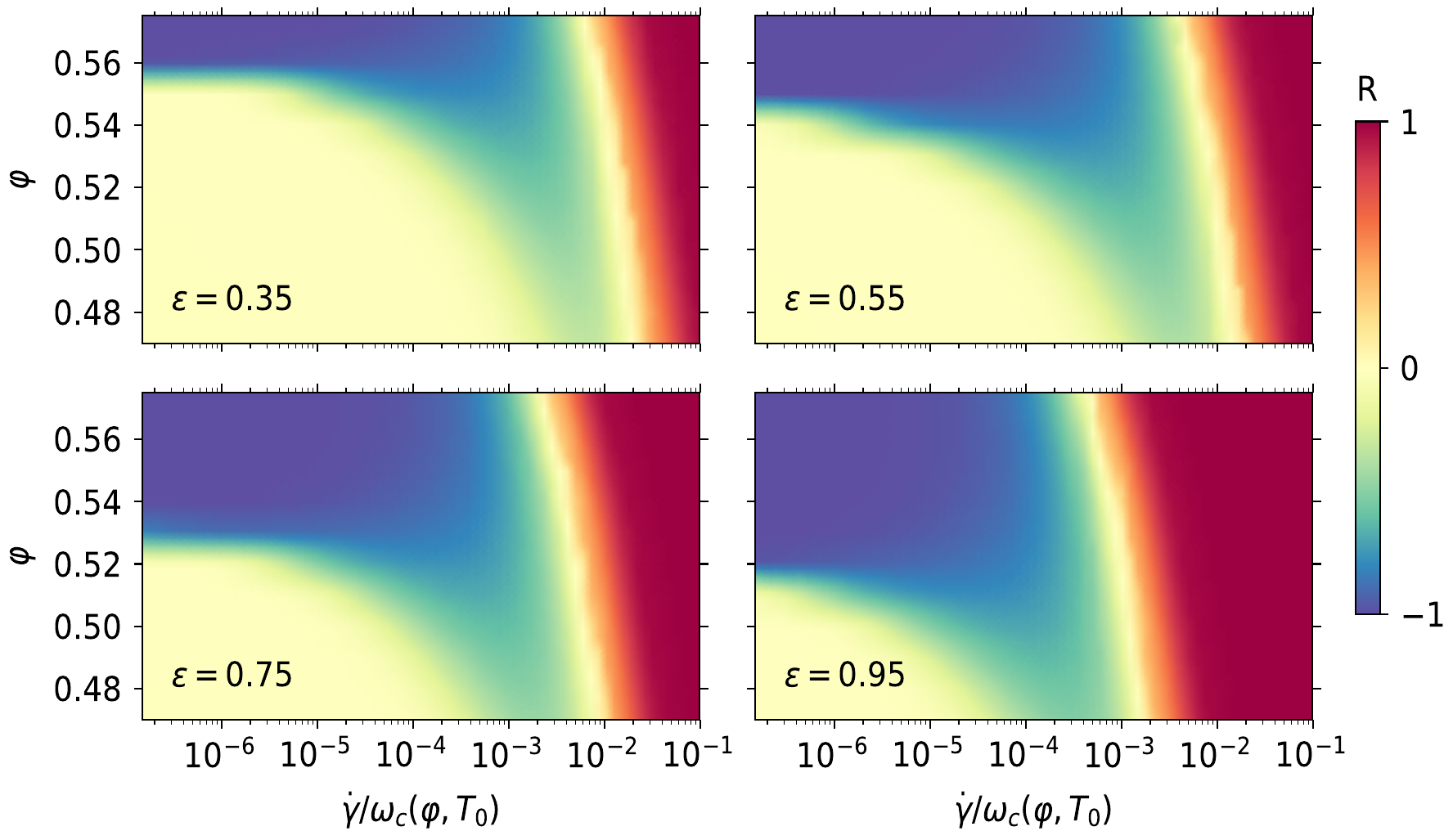}
  \caption{\textbf{Protocol H} (Shear heating increases the granular
    temperature): Dynamic state diagrams in the plane spanned by shear
    rate, $\dot\gamma$, and packing fraction, $\varphi$, for several
    values of the coefficient of restitution, $\varepsilon$, as
    indicated. The effective exponent $R$ quantifying the shear rate
    dependence of the viscosity,
    $\eta(\dot\gamma) \sim {\dot\gamma}^R$, is color coded. Shear
    thickening (thinning) corresponds to $R > 0$ ($R<0$) while $R=0$
    indicates Newtonian behavior. The shear rate is normalized by the
    collision frequency $\omega_c(\varphi, T_0)$ of the unsheared
    system (at granular temperature $T_0$) with the prescribed volume
    fraction. The data for all figures is available as Supplemental
    Material \cite{suppnote}.}
  \label{fig:regimes}
\end{figure}

Qualitative considerations have shown that the dynamic state diagram
characterizing the rheological response of a granular fluid to shear
(Fig.~\ref{fig:regimes}) is determined by several timescales
\cite{kranz+frahsa17a}. The Newtonian behavior---where shear stress
$\sigma = \eta\dot\gamma$ is linearly related to the shear rate
$\dot\gamma$---extends to higher densities and shear rates as long as
the shear rate is smaller than the intrinsic structural relaxation
rate $\tau_{\alpha}^{-1}$. Beyond that, the shear stress becomes
approximately independent of shear rate, indicated by shear thinning
behavior. For even higher shear rates, shear heating will become
important which appears as shear thickening behavior. Ultimately,
Bagnold scaling, $\sigma \sim {\dot\gamma}^2$, holds for the largest
shear rates.

In this contribution we will introduce a kinetic theory, namely the
granular Integration Through Transients (\textsc{gitt}) formalism that
is tailored to high densities and arbitrary shear rates of an
intrinsically out-of-equilibrium granular fluid. To be precise,
\textsc{gitt} is expected to be most accurate close to the granular
glass transition
\cite{abate+durian06,kranz+sperl10,gholami+fiege11,avila+castillo16},
\textit{i.e.}, for packing fractions $\varphi$ in the range
$0.5\lesssim\varphi\lesssim0.6$. In addition, \textsc{gitt} is
\emph{not} an expansion around zero shear and also \emph{not} limited
to almost elastic particles. Our approach is based on the \textsc{itt}
formalism for thermalized colloidal dispersions
\cite{fuchs+cates02,fuchs+cates09,chong+kim09,brader+voigtmann09,nicolas+fuchs16}
and provides a framework to calculate flow curves, \textit{i.e.},
shear stress $\sigma$ as a function of shear rate $\dot\gamma$, beyond
the linear response regime. To keep the paper at a manageable length
we will use the simplest assumptions and point out possible
refinements for future work.

A first generalization of the \textsc{itt} formalism from the
overdamped Brownian dynamics of colloidal suspensions to inertial,
Newtonian dynamics has been obtained by \citet{chong+kim09} and later
updated to the refined formulation of \citet{fuchs+cates09} by
\citet{suzuki+hayakawa13}. In both cases the lack of viscous damping
in the inertial description necessitates the explicit introduction of
a thermostat. At least for the Gaussian isokinetic thermostat
\cite{evans+morriss07} used in the above references, the details of
the artificial thermostat explicitly appear in the final results.
\citet{suzuki+hayakawa14} were the first employing an \textsc{itt}
calculation for inelastic \emph{soft} spheres. Unexpectedly, they found
that no dissipative effects remain after applying the standard
\textsc{itt} approximations. Only after including current correlation
functions in addition to the density correlator did they find results
that depend on the inelasticity of the particles. The substantially
increased complexity due to the additional observables has so far made
applications of this approach difficult. On the upside, employing
dissipative interactions removes the need for an artificial
thermostat.

In the following we will show that using frictionless inelastic
\emph{hard} spheres, it is much simpler to retain dissipative effects
and this allows us to develop a granular \textsc{itt} formalism that
captures the qualitative behavior introduced above.  The paper is
organized as follows. We will start with an informal outline of the
model and derivations in Sec.~\ref{sec:outl-deriv-main}. Subsequently
in Sec.~\ref{microscopics}, we introduce the microscopic equations of
motion and the observables, i.e., correlation functions. The
generalized Green-Kubo relations that form the central part of the
granular \textsc{itt} formalism (\textsc{gitt}) will be derived in
Sec.~\ref{sec:green-kubo}. To evaluate the Green-Kubo relations, we
need to compute the transient correlator approximately in mode-coupling
theory, as discussed in Sec.~\ref{sec:mode-coupling-theory}. In
Sec.~\ref{sec:results} we will present a number of results that can be
derived from the formalism developed here before concluding in
Sec.~\ref{sec:conclusion}. 

\section{Outline of the Derivation and Main Assumptions}
\label{sec:outl-deriv-main}

As the following derivation comprises a number of technical steps,
requiring a set of assumptions and approximations, we are going to
outline the main steps of the derivation here informally, pointing to
the relevant details in the later sections. We restrict our analysis
to the simplest model that captures the essential physics and relegate
the discussion of some exemplary extensions to
Sec.~\ref{sec:conclusion}.

Starting from a quiescent granular fluid, beginning at time $t=0$ we
impose a shear profile of the form
$\vec u(\vec r) = \vec r\cdot\mathsf{k}$ with a velocity gradient
tensor $\mathsf{k}$ which is given by
$k_{\alpha\beta} = \dot\gamma\delta_{\alpha x}\delta_{\beta y}$.  To
compute the shear stress $\sigma(\dot\gamma)$, we work along the
following roadmap:
\begin{enumerate}
  \renewcommand{\theenumi}{\Alph{enumi}}
\item Specify the stationary state of the quiescent fluid
\item Specify the stationary state of the sheared fluid which is reached for asymptotically long times
\item Generalize the Green-Kubo relation and integration through
  transients to granular fluids
\item Couple stress and structural relaxation
\item Compute the transient correlations within mode-coupling approximation
\end{enumerate}

\subsection{The Quiescent State}
\label{sec:flow-problem}

We consider a granular fluid comprised of $N$ monodisperse
frictionless inelastic hard spheres of diameter $d$, mass $m = 1$ and
density $n=N/V$. The inelasticity is modeled by a constant coefficient
of normal restitution $\varepsilon\in[0,1)$
\cite{haff83,aspelmeier+huthmann01}. Note that the incomplete
restitution in collision breaks time reversal symmetry. We assume that
despite the monodispersity the system will not crystallize at any
time. This choice is motivated by the observation that a small amount
of polydispersity, just enough to prevent crystallization
\cite{scheope+bryant07}, does not qualitatively alter the behavior of
hard sphere fluids but only yields small quantitative corrections
\cite{hunter+weeks12}. Considering the monodisperse limit simplifies
the calculations, though. Note, however, that the quiescent fluid
undergoes a granular glass transition to an amorphous solid above a
critical packing fraction $\varphi_c(\varepsilon)$
\cite{abate+durian06,kranz+sperl10,sperl+kranz12,kranz+sperl13}. Away
from the jamming transition, macroscopic granular particles are well
approximated by hard-core interactions. Considerable particle
deformations are expected to become relevant in violent flow
conditions with very high impact speeds which consequently fall
outside of our model.

The only intrinsic length scale in the system is the particle diameter
$d$ and the only energy scale is given by the granular temperature $T$
(we set the Boltzmann constant $k_B=1$ throughout). The quiescent
fluid therefore comprises a single time scale that may be given as the
inverse of the collision frequency $\omega_c \propto \sqrt T/d$
\cite{chapman+cowling70}. The dimensionless measure of density is the
packing fraction $\varphi := \pi nd^3/6$.

Considering molecular or colloidal fluids it is natural to assume that
shear is applied to a quiescent state in thermal equilibrium. For a
granular system, there is no such uniquely defined quiescent fluid.
In order to have a stationary granular fluid even before shear is
applied, we imagine the granular system to be under the influence of a
homogeneous driving force. Realization of such a driving force include
fluidization by air
\cite{ojha+lemieux04,taghipour+ellis05,born+schmitz17} or water
\cite{schroeter+goldman05}.  To concentrate on the granular aspects of
the rheology, we do not attempt to model the driving mechanism in any
detail. Instead we assume that it results in a homogeneous random
driving force that injects a prescribed amount of power, $P_D$, into
the system to compensate the energy dissipated in inelastic collisions
\cite{williams+mackintosh96,kranz+sperl10,fiege+aspelmeier09}. The
granular temperature, $T_0$, of the quiescent fluid is then implicitly
given by the power balance \cite{brilliantov+poeschel10,pagonabarraga+trizac01,garzo+montanero02,fiege+aspelmeier09,kranz+sperl10},
\begin{equation}
  \label{eq:1}
  P_D = \Gamma\omega_c(T_0)T_0,
\end{equation}
between driving and dissipation. Here
$\Gamma = \Gamma(\varphi,\varepsilon)$ is nothing but the
dimensionless dissipation rate \footnote{In Enskog approximation we
  have
  $\Gamma(\varphi,\varepsilon) \equiv \Gamma_E(\varepsilon) := (1 -
  \varepsilon^2)/3$ independent of density \cite{haff83}.}.

The \textsc{itt} formalism, discussed below, allows to express
ensemble averages in the strongly sheared stationary state as
expectation values with respect to the phase space distribution of a
quiescent reference fluid, $\varrho_{\mathrm{ref}}$.  In thermal
equilibrium, the quiescent fluid is described by the canonical
distribution. Currently, a theoretical framework to determine
$\varrho_{\mathrm{ref}}$, comparable to the framework of statistical
ensembles in thermal equilibrium, does not exist. Only the marginal
one-particle velocity distribution function $\varrho_v^{(1)}(\vec c)$
is well characterized \cite{brilliantov+poeschel10}. It is
non-Gaussian but with rapidly decaying tails,
$\ln\varrho_v^{(1)}(c \to \infty) \propto -c^{3/2}$ and its dominant
contribution can be expressed as an expansion in Sonine polynomials
around a Gaussian \cite{brilliantov+poeschel10}. The exponentially
decaying tails guarantee that the second moment of the velocity
distribution exists and the quiescent reference fluid displays a well
defined granular temperature:
\begin{equation}
  \label{eq:2}
  T_0 := \frac13\avr{\vec c^2},
\end{equation}
where the angular brackets denote the phase space average with respect
to $\varrho_{\mathrm{ref}}$.

The most severe approximation we make is to factorize the phase space
distribution into a positional and a velocity part
$\varrho_{\mathrm{ref}} \approx \varrho_r^{(N)}(\vec r_1,\ldots,\vec
r_N)\varrho_v^{(N)}(\vec c_1,\ldots, \vec c_N)$ although correlations
are known to exist \cite{pagonabarraga+trizac01}. Here $\vec r_i$
denotes the particles' positions and $\vec c_i$ their peculiar
velocities [cf.~Eq.~(\ref{eq:14})]. Let us stress that in the
\textsc{itt} formalism we only need an explicit expression for
$\varrho_{\mathrm{ref}}$ in the unsheared state. The strong anisotropy
of the distribution of relative collisional velocities $\vec c_{ij}$
with respect to the impact parameter in the sheared stationary state
of a granular fluid \cite{kumaran09b}, are---although an interesting
observation---unrelated to our factorization approximation.  We
furthermore assume the precollisional velocity distribution to
factorize in particles,
\begin{equation}
  \label{eq:3}
  \varrho_v^{(N)}(\vec c_1,\ldots,\vec c_N) =
  \prod_{i=1}^N\varrho_v^{(1)}(\vec c_i).
\end{equation}
While the postcollisional velocities are certainly highly correlated,
this is a less severe approximation for the precollisional velocities
\cite{lutsko96,lutsko01}. Moreover we effectively assume a Gaussian
form,
$\varrho_v^{(1)}(\vec c) \propto \exp(-\vec c^2/2T)$.
Note, however, that in many places we only need the second moment of
the distribution regardless of its overall form. 

The spatial structure of a dense granular fluid has received
comparably little attention. A structure factor theory for granular
fluids has not been developed to the best of our
knowledge. Simulations and experiments in two dimensions
\cite{puglisi+gnoli12} indicate relatively modest deviations from the
equilibrium hard sphere structure. Keeping to the simplest
approximations, we assume the spatial structure in the granular fluid
to be identical to the structure of an elastic hard sphere fluid in
thermal equilibrium,
$\varrho_r^{(N)}(\vec r_1,\ldots,\vec r_N) \propto
\prod_{j<k}\Theta(r_{jk}-d)$.
In summary, we effectively assume the phase space distribution (but
not the dynamics!) in the quiescent fluid to be of canonical form.

\subsection{Stationary State of the Sheared Fluid}
\label{sec:stat-state-shear}

In contrast to colloidal suspensions where the particles are embedded
in a host fluid that acts as a heat bath, the granular fluid is not
necessarily thermostated. Once shear is applied, the work expended to
sustain the shear flow injects additional energy into the fluid. In
colloidal suspensions under normal conditions, this additional shear
heating, $\sigma\dot\gamma$ (naturally defined per volume), is readily
absorbed by the heat bath and the thermodynamic temperature of the
suspensions remains constant. For the granular fluid we are going to
consider two protocols that deal in different ways with the additional
shear heating \footnote{Let us stress here that by \emph{shear
    heating} of a \emph{granular fluid}, we mean the transfer of the
  work required to shear the fluid into a state with finite kinetic
  energy of the granular particles, \textit{i.e.}, into a finite
  granular temperature.}:
\begin{description}
\item[Protocol H] keeping the driving power, $P_D$, fixed between the
  quiescent and the sheared fluid such that the shear heating
  increases the total injected power,
  $P_D + \sigma\dot\gamma/n =\Gamma\omega_c(T)T$. Using
  Eq.~(\ref{eq:1}) to eliminate $P_D$, the power balance may be
  rewritten as
  \begin{equation}
    \label{eq:4}
    \sigma\dot\gamma + n\Gamma\omega_c(T_0)T_0 = n\Gamma\omega_c(T)T.
  \end{equation}
  The granular temperature in the sheared stationary state $T > T_0$
  will necessarily be elevated compared to the quiescent fluid.
\item[Protocol T] keeping the total injected power fixed, \textit{i.e.},
  reducing the driving power in the sheared fluid
  $P_D^{\dot\gamma} < P_D$, such that
  $P_D^{\dot\gamma} + \sigma\dot\gamma/n = P_D$. Note that this
  effective thermostatting can only work as long as the shear heating
  is smaller than the initial driving power,
  $\sigma\dot\gamma \leq nP_D$, \textit{i.e.}, it only applies up to a
  maximum shear rate $\dot\gamma_{\infty}$. For admissible shear
  rates, however, the granular temperature in the sheared stationary
  state $T \equiv T_0$ will remain unchanged from the quiescent fluid.
\end{description}
Note that \textsc{gitt} formalism developed below is not limited to
these simple protocols. More complicated power balances can easily be
incorporated.

The time scale set by the shear rate $\dot\gamma$ is to be compared to
the intrinsic time scale, defining the dimensionless P\'eclet number
$\Pe := \dot\gamma/\omega_c(T)$.  Besides the coefficient of
restitution $\varepsilon$, the only two parameters which control the
rheology \cite{kranz+frahsa17a} are $\varphi$ and $\Pe$. A
dimensionless shear stress is naturally defined as
$\hat\sigma = \hat\sigma(\Pe; \varphi, \varepsilon) := \sigma/nT$,
independent of the granular temperature $T$.
At low density, $\varphi\ll1$, and small shear rate, $\dot\gamma\to0$,
we expect Newtonian rheology \cite{kranz+frahsa17a},
\begin{equation}
  \label{eq:5}
  \sigma(\varphi\ll1) =: \sigma_0 = \eta\dot\gamma,
\end{equation}
with a viscosity $\eta(\varphi\ll1)$ that is well approximated by the
Enskog predictions, $\eta_E(\varphi, \varepsilon)$, of
\citet{garzo+montanero02}.

Solving Eq.~(\ref{eq:4}) for the granular temperature $T$ in the
sheared stationary state of \emph{Protocol H}, one finds
\footnote{Note that a formula for $T(\dot\gamma)$ cannot be given in
  closed form}
\begin{equation}
  \label{eq:6}
  T(\Pe) = \frac{T_0}{(1 - \Pe/\Pe_{\infty})^{2/3}},
\end{equation}
where $\Pe_{\infty}$ is the maximum admissible P\'eclet number
implicitly defined by \cite{kranz+frahsa17a}
\begin{equation}
  \label{eq:7}
  \Pe_{\infty} = \Gamma/\hat\sigma(\Pe_{\infty}).
\end{equation}
Note that the stationary granular temperature $T$ diverges as
$\Pe\to\Pe_{\infty}$, however, contrary to \emph{Protocol T}, this
does \emph{not} restrict the shear rate in \emph{Protocol H},
$\dot\gamma = \omega_c(T)\Pe$, as the collision frequency diverges
together with the granular temperature.  

\subsection{Generalized Green-Kubo Relation}
\label{sec:gener-green-kubo}
At times $t < 0$ the fluid is assumed to be in the quiescent reference
state at the granular temperature $T$ maintained by an associated
driving power $P_D^{\mathrm{ref}} = \Gamma\omega_c(T)T$ according to
Eq.~(\ref{eq:1}). At the time $t = 0^+$ shear is switched on
instantaneously and subsequently the system is given a long---formally
infinite---time to relax through a transient phase into the new
sheared stationary state, characterized by $ \varrho_{\mathrm{ss}}$. We do not control the transient phase in any way. In fact, it is to be
expected that the granular temperature will be time dependent in the
transient regime. To keep the temperature variations as small as
possible, we assume the granular temperature of the reference state to
be identical to the granular temperature in the sheared stationary
state, $T$. Note that in \emph{Protocol H}---where $T > T_0$---this
implies that the reference state is kept at a higher granular
temperature than the quiescent unsheared fluid. Only in \emph{Protocol
  T}, all granular temperatures are equal, $T = T_0$.  This implies a
reduction of the driving power from $P_D^{\mathrm{ref}}$ to
$P_D^{\dot\gamma} \leq P_D < P_D^{\mathrm{ref}}$.

The starting point of the \textsc{itt} formalism
\cite{fuchs+cates02,fuchs+cates09} is the identity
\begin{equation}
  \label{eq:8}
  \varrho_{\mathrm{ss}} = \varrho_{\mathrm{ref}} 
  + \int_0^{\infty}dt'\frac{d\varrho_{\mathrm{ref}}}{dt}(t')
\end{equation}
expressing the phase space distribution function in the sheared
stationary state, $\varrho_{ss}$, in terms of the distribution
function of the unsheared system, $\varrho_{\mathrm{ref}}$. The
evolution of $\varrho_{\mathrm{ref}}$ into $\varrho_{\mathrm{ss}}$ is
then calculated by an integration through the eponymous
\emph{transient} in time. As the reference distribution is stationary
under the quiescent dynamics, only the shear part of the dynamics
contributes to the time derivative. With the assumption of a canonical
$\varrho_{\mathrm{ref}} \propto\exp(-E_{\mathrm{int}}/T)$, written in
terms of the internal energy of the fluid $E_{\mathrm{int}}$
(cf.~appendix~\ref{sec:sllod-equat-moti}), one finds
\cite{fuchs+cates09,chong+kim09}
\begin{equation}
  \label{eq:9}
  \frac{d\varrho_{\mathrm{ref}}}{dt}(t')
  = -\dot\gamma\sigma_{xy}^{\mathrm{el}}(t')\varrho_{\mathrm{ref}}/T.
\end{equation}
Here $\sigma_{xy}^{\mathrm{el}}$ denotes the Irving-Kirkwood
expression for the shear stress of elastic hard spheres in terms of
the phase space variables \cite{dufty12}.

The macroscopic shear stress
$\sigma \equiv -\avr{\sigma_{xy}}_{\mathrm{ss}}/V$ in the sheared
stationary state may then be evaluated in terms of a generalised
Green-Kubo integral (cf.~Sec.~\ref{sec:green-kubo})
\begin{equation}
  \label{eq:10}
  \sigma
  = \frac{\dot\gamma}{VT}
  \int_0^{\infty}dt
  \avr{\sigma_{xy}(t=0)\mid\sigma_{xy}^{\mathrm{el}}(t)},
\end{equation}
expressed in terms of a correlation function in the unsheared
reference state. The microscopic granular shear stress
$\sigma_{xy} = \sigma_{xy}^{\mathrm{kin}} +
\sigma_{xy}^{\mathrm{int}}$
is composed of a kinetic part, $\sigma_{xy}^{\mathrm{kin}}$, and a
collisional part, $\sigma_{xy}^{\mathrm{int}}$
(cf. appendix~\ref{sec:density-stree}). The generalised Green-Kubo
relation (\ref{eq:10}) is formally identical to the colloidal results
\cite{fuchs+cates02,fuchs+cates09}.  Note that the reference state
does not carry a shear stress $\avr{\sigma_{xy}} \equiv 0$ and that
the microscopic collisional stress,
$\sigma_{xy}^{\mathrm{int}} = \frac{1 +
  \varepsilon}{2}\sigma_{xy}^{\mathrm{int,el}}$,
Eq.~(\ref{eq:79}), in a configuration of granular particles is lower
than the stress of the identical configuration of elastic particles
due to the incomplete restitution of momentum upon collision
\cite{baskaran+dufty08}. Furthermore, the
dynamics of the stress auto-correlation function
$\avr{\sigma_{xy}\mid\sigma_{xy}(t)}$ will be affected by the granular
dynamics and its broken time-reversal symmetry

Note that we have lost all reference to the distribution function in
the sheared stationary state, $\varrho_{ss}$. Instead, the effect of
shear has been transfered to the dynamics $\sigma_{xy}(t)$ in the
transient buildup of the stationary state. This is fundamentally
different from, say, the Chapmann-Enskog procedure where one attempts
to describe the phase space distribution function in the sheared
stationary state perturbatively \cite{chapman+cowling70}. Let us
stress that in deriving Eq.~(\ref{eq:10}) at no place have we assumed
that the shear rate or the shear stress are small. Consequently, the
generalized Green-Kubo relation is not an expansion in small
perturbations of the reference state. For colloids ($\varepsilon=1$)
it is exact and its validity for granular fluids is only restricted by
(i) the degree to which Eq.~(\ref{eq:9}) is fulfilled, and (ii) the
magnitude of the residual effects of the change in driving power if
applicable. We stress, however, that no entropy arguments have been
employed that are \textit{a priori} restricted to thermal
equilibrium. This is to be contrasted with the formally very similar
Green-Kubo relations derived by Ronis that are based on entropy
maximization and constitute but the lowest order in an expansion in
$\dot\gamma$ \cite{ronis79,croteau+ronis02}. Consequently, the
Green-Kubo relation~(\ref{eq:10}), although it contains an explicit
prefactor $\dot\gamma$, is able to capture non-linear constitutive
laws including shear thinning and shear thickening
\cite{fuchs+cates02,kranz+frahsa17a}. Applying Ronis' approach to a
homogeneously cooling granular fluid \cite{goldhirsch+vannoije00},
instead, captures Newtonian rheology only.

\subsection{Coupling of Stress and Structural Relaxation}
\label{sec:struct-relax}

The time evolution of the transient stress auto-correlation function
is not known. However, in dense fluids we may assume it to be enslaved
to the slow structural relaxation encoded in the glassy dynamics of
the density fluctuations \cite{fuchs+cates02,fuchs+cates09}. Note,
however, that the approximation introduced below does not capture the
jamming transition. This implies that densities are restricted to be
sufficiently far below the random close packing density
$\varphi_{\mathrm{rcp}}\approx0.64$ \cite{aste+weaire00} not to be
dominated by the imminent jamming transition. The slow relaxation is
intrinsically a collective phenomenon that occurs on macroscopic
scales \cite{debenedetti+stillinger01}. To this end we may adopt a
coarse grained, continuum description in terms of a continuous density
field $\rho(\vec r)$. To maintain a continuum description, we need to
make sure that we do not impose excessive velocity gradients,
\textit{i.e.}, we need to require that the Knudsen number $\Kn < 1$.
Here we can take the P\'eclet number as a proxy for the Knudsen number
\footnote{The Knudsen number $\Kn = \ell_0/L$ is the dimensionless
  ratio of the mean free path, $\ell_0$, to a characteristic
  length-scale for gradients, $L$. Defining $L$ as the length over
  which the shear speed becomes comparable to the thermal velocity,
  $\dot\gamma L \sim \sqrt T\sim\omega_c\ell_0$, we have
  $\Kn\sim\Pe$.} and the fact that the P\'eclet number is limited
$\Pe \leq \Pe_{\infty}$ \cite{kranz+frahsa17a} ensures that we are not
limited in shear rates as long as the density is sufficiently high
such that $\Pe_{\infty} < 1$.

It has been noted that shear flow distorts plane waves
\cite{fuchs+cates02,kumaran09c}, so that translational invariance
holds only in the comoving frame \cite{fuchs+cates09}. Therefore, we
define density correlations
\begin{equation}
  \label{eq:11}
  \Phi_{\vec k}(t', t) := N\avr{\rho_{\vec k}(t')\mid\rho_{\vec
      k(t - t')}(t)}/S_{\vec k}(t')
\end{equation}
as the overlap of a density fluctuation $\rho_{\vec k}$ with wave
vector $\vec k$ at time $t'$ with another density fluctuation at a
later time $t > t^{\prime}$ with the advected wave vector
$\vec k(t - t') = \vec k\cdot[\mathsf{1} - \mathsf{k}(t - t')]$.  The
averages are evaluated with the time-independent distribution
$\varrho_{\mathrm{ref}}$ of the reference state. However, the full
time evolution, including shear, is not stationary in the transient
regime. Therefore, the transient correlators in general depend on two
time arguments, while the equal time correlation
$S_{\vec k}(t') := N\avr{\rho_{\vec k}(t')\mid\rho_{\vec k}(t')}$
depends on the waiting time $t'$. It turns out we are only ever going
to make use of a zero waiting time, $t' = 0$, and simplify notation to
$\Phi_{\vec k}(t) \equiv \Phi_{\vec k}(0, t)$. Incidentally, the zero
waiting time structure factor $S_{\vec k}(0) \equiv S_{\vec k}$ is
identical to the structure factor in the quiescent reference state.

In comparing wave vectors at time $t$ and time $t=0$ one may
equivalently regard either point in time as the one defining wave
vector $\vec k$ of interest and then determine the advected wave
vector $\vec k(\pm t)$. While fixing the wave vectors at $t = 0$ to
be advected to time $t$ seems natural \cite{fuchs+cates02}, it was
later found to be advantageous to fix the wave vector at time $t$
and relate it to the wave vector $\vec k(-t)$ at time $t=0$ through
backwards advection \cite{fuchs+cates09,suzuki+hayakawa13}.

Momentum conservation dictates that the stress tensor couples to pairs
of density modes, $\rho_{\vec k}\rho_{-\vec k}$. The time evolution of
the pair modes is determined by a four-point density correlator, which
is approximately factorized into products of two-point density
correlators.  By way of this approximation we have transferred the
generalized Green-Kubo relation from its natural formulation in terms
of the stress auto-correlation function, Eq.~(\ref{eq:10}), to a
formulation, albeit approximate, in terms of the density
auto-correlation [cf.~Eq.~(\ref{eq:33})]
\begin{equation}
  \label{eq:12}
  \sigma = \sigma_0 + \frac{1 + \varepsilon}{4}\times\frac{\dot\gamma}{T}
  \int\frac{d^3k}{(2\pi)^3}\int_0^{\infty}dt
  \frac{\mathcal V_{\vec k(-t)}^{\sigma}\mathcal V_{\vec k}^{\sigma}}{S_k^2}
  \Phi^2_{\vec k(-t)}(t).
\end{equation}
The coupling between stress and density is given by anisotropic, wave
number dependent coupling constants $ \mathcal V_{\vec k}^{\sigma} $,
which will be derived explicitly in Sec.~\ref{sec:green-kubo}. Note
that the transient correlator, $\Phi_{\vec k(-t)}(t)$, implicitly
depends on the coefficient of restitution $\varepsilon$ and the shear
rate $\dot\gamma$. However, we also retain the explicit dependence of
the shear stress $\sigma$ on the coefficient of restitution
$\varepsilon$, as a result of the non-linearity of the collision
rule. If we had modeled the dissipative collisions by a linear viscous
law, we would have lost all dissipative effects at this point. For
details see appendix~\ref{sec:case-visc-diss}. Note that due to the
approximation in terms of density pairs, the generalized Green-Kubo
integral is of the order $\varphi^2$. In order to extend the validity
of Eq.~(\ref{eq:12}) to lower densities, we have added the low density
contribution $\sigma_0$, Eq.~(\ref{eq:5}).

The spatial structure of the isotropic quiescent reference
fluid---partially encoded in the static structure factor $S_k$---is,
in general, also a function of the coefficient of restitution
$\varepsilon$. Note that the anisotropy of the sheared state is
encoded in two ways in Eq.~(\ref{eq:12}): (i) by the anisotropic form
of the coupling constant, Eq.~(\ref{eq:32}), and (ii) by the
anisotropy of the advected wave vector $\vec k(-t)$. By virtue of the
\textsc{itt} approach, however, it is sufficient to consider the
spatial structure of the isotropic reference state. This is to be
contrasted with theoretical approaches that approximate the
distribution of the sheared stationary state, $\varrho_{\mathrm{ss}}$,
more directly, including its spatial anisotropy
\cite{kumaran09a,kumaran09b}.

\subsection{Transient density correlations}
\label{sec:trans-dens-corr}

The dynamics of the transient density correlator may be evaluated
self-consistently in the framework of mode-coupling theory. The
derivation of the equation of motion for $\Phi_{\vec k}(t)$ is quite
technical and constitutes a major part of this contribution
(cf.~Sec.~\ref{sec:mode-coupling-theory}). In essence, we derive the
linearized, compressible, fluctuating hydrodynamic equations taking
into account the wave vector advection due to shear.

The first step is to take care of the conservation laws of particle
density and momentum with help of the Mori-Zwanzig projection operator
formalism, following closely \citet{suzuki+hayakawa13}. The next step
is a mode-coupling approximation for the dynamics of the resulting
memory kernel, assuming that the decay is dominated by pairs of
density fluctuations. Thereby the memory kernel is expressed in terms
of four-point density correlations, which are subsequently
factorized. We end up with a set of self-consistent equations for the
time-dependent, transient correlations of particle densities and currents.

As compared to mode-coupling theories in a quiescent state, an
additional complication arises, because the applied shear breaks the
isotropy of the fluid. Solving the full set of equations is
challenging even numerically \cite{amann+denisov15} and as a first
step we are not going to attempt it. Previous studies of glass forming
colloidal dispersions have shown that density correlations remain
comparably isotropic \cite{varnik+henrich06,besseling+weeks07}.
Importantly, this does not rule out finite shear elements of the
stress tensor, which are encoded in the mode-coupling
vertices. \Citet{chong+kim09} have argued that the essential
anisotropy in Eq.~(\ref{eq:12}) is encoded in the coupling constants
$\mathcal V_{\vec k}^{\sigma}$ such that it is sufficient to consider
an isotropic approximation for the transient correlator $\Phi_k(t)$
that only depends on the modulus but not on the direction of the wave
vector. The resulting equations are solved numerically by iteration.

\section{Microscopic Dynamics}
\label{microscopics}
\subsection{Pseudo Liouville operator}
For the implementation of the imposed linear shear
profile, $\vec u(\vec r) = \vec r\cdot\mathsf{k}$ we use the Sllod equations
\cite{evans+morriss07}. We sketch their derivation for Hamiltonian
systems in appendix~\ref{sec:sllod-equat-moti}. In terms of the
dissipative hard sphere interactions and for the given shear profile
they provide the equations of motion for the particles' positions,
$\vec r_i(t)$, and peculiar velocities, $\vec c_i(t)$,
\begin{subequations}
  \label{eq:13}
\begin{align}
  \label{eq:14}
  \dot{\vec r}_i &= \vec c_i + \vec r_i\cdot\mathsf{k}\\
  \label{eq:15}
  \dot{\vec c}_i &= \left.\frac{d\vec c_i}{dt}\right|_{\mathrm{coll}}
                 + \vec\xi_i(t) - \vec c_i\cdot\mathsf{k}.
\end{align}
\end{subequations}
Here, $d\vec c_i/dt|_{\mathrm{coll}}$ denotes the effect of
dissipative hard sphere collisions and we have used the property
$\mathsf{k}\cdot\mathsf{k}\equiv\mathsf{0}$. Eq.~(\ref{eq:14})
implicitly defines the peculiar velocity $\vec c_i$, whereas
Eq.~(\ref{eq:15}) describes that the peculiar velocity may change for
one of three reasons: Firstly, due to collisions with other particles,
secondly, due to the random driving force, and, finally, if the
particle moves in gradient direction, the flow velocity with respect
to which its peculiar velocity is defined changes, effectively
changing the peculiar velocity. The random force $\vec\xi_i(t)$
\cite{williams+mackintosh96,bizon+shattuck99} has zero mean
$\overline{\vec\xi_i} = 0$ and variance
\begin{equation}
  \label{eq:16}
  \overline{\xi_i^{\alpha}(t)\xi_j^{\beta}(t')} 
  = \frac23P_D\delta_{ij}\delta^{\alpha\beta}\delta(t-t').
\end{equation}
Note that in the unsheared reference system, the peculiar velocities
are identical to the lab-frame velocities,
$\vec v_i = \dot{\vec r}_i\equiv \vec c_i$.

The above dynamics can be rephrased in terms of a (forward in time)
effective pseudo Liouville operator $\Lv_+$ that generates the time
derivative of arbitrary functions $A(X)$ on the phase space $X$,
$dA/dt \equiv \Lv_+A$. In the unsheared reference system the dynamics
\cite{kranz+sperl13}
\begin{equation}
 \Lv^+_{\mathrm{ref}} = \Lv_0 +\Lv_I^+ + \Lv_D^+(P_D)
\end{equation}
consists of free streaming, $i\Lv_0 = \sum_j\vec c_j\cdot\nabla_j$,
random driving, $\Lv_D^+(P_D) = P_D\sum_j\partial^2_{\vec c_j}$, and
binary collisions $i\Lv_I^+ = \sum_{j<k}i\mathcal T_{jk}^+$, where
\begin{equation}
  \label{eq:17}
  i\mathcal T_{jk}^+ = -(\uvec r_{jk}\cdot\vec v_{jk})
  \Theta(-\uvec r_{jk}\cdot\vec v_{jk})\delta(r_{jk} -
  d)(\mathsf{b}_{jk}^+ - \mathsf{1})
\end{equation}
denotes the inelastic binary collision operator
\cite{vannoije+brito98,aspelmeier+huthmann01} that explicitly breaks
time-reversal symmetry. Here the operator $\mathsf{b}_{jk}^+$
implements the inelastic collision rule \cite{huthmann+zippelius97},
$\vec r_{jk} := \vec r_j - \vec r_j$, likewise for $\vec v_{jk}$, and
$\Theta(x)$ denotes the Heaviside step function. 

The Sllod equations give rise to an additional contribution to the
Liouville operator
$\Lv_{\dot\gamma} = \Lv_{\dot\gamma_r} + \Lv_{\dot\gamma_c}$
\cite{chong+kim09}, with
\begin{equation}
  \label{eq:18}
  i\Lv_{\dot\gamma_r} = \sum_j\vec r_j\cdot\mathsf{k}^{\mathsf t}\cdot\nabla_j,\quad
  i\Lv_{\dot\gamma_c} = -\sum_j\vec c_j\cdot\mathsf{k}^{\mathsf t}\cdot
  \frac{\partial}{\partial\vec c_j}.
\end{equation}
such the full dynamics in the sheared state is generated by
$\Lv_+ = \Lv^+_{\mathrm{ref}} + \Lv_{\dot\gamma}$

Note that in contrast to all the other parts of the Liouville
operator, the binary collision operator, Eq.~(\ref{eq:17}), is
formulated in terms of the lab-frame velocities $\vec v_i$ as opposed
to the peculiar velocities $\vec c_i$. By virtue of the generalized
Green-Kubo relation (\ref{eq:10}), however, all matrix elements of the
binary collision operator in the following will be evaluated in the
unsheared reference state where the two frames of reference
coincide. In the calculation of the stress tensor within the
Chapman-Enskog approach \cite{chapman+cowling70}, the mismatch between
the $\vec v_i$ and $\vec c_i$ is crucial for the collisional part of
the stress induced by the particles' finite size. In the \textsc{itt}
approach, the finite size of particles is encoded in a non-trivial
structure factor $S_k$.

\subsection{Ensemble averages in the reference system}
The ensemble average of an observable $A(X)$ in the reference system
is defined as a scalar product
\begin{equation}
  \label{eq:19}
  \frac{d\avr A}{dt} = \avr{i\Lv^+_{\mathrm{ref}}A}
  = \int dX\varrho_{\mathrm{ref}}(X)\Lv^+_{\mathrm{ref}}A(X)
  =: (\varrho_{\mathrm{ref}}, \Lv^+_{\mathrm{ref}}A)
\end{equation}
allowing us to introduce the Liouvillean $\overline\Lv^+_{\mathrm{ref}}$ as
the adjoint of $\Lv^+_{\mathrm{ref}}$,
\begin{equation}
  \label{eq:20}
  (\overline\Lv^+_{\mathrm{ref}}\varrho_{\mathrm{ref}}, A) = (\varrho_{\mathrm{ref}}, \Lv^+_{\mathrm{ref}}A),
\end{equation}
acting on the phase space distribution $\varrho_{\mathrm{ref}}$. With respect
to two observables $A_1(X),A_2(X)$, we write correlation functions as another
scalar product,
\begin{equation}
  \label{eq:21}
  \avr{A_1\mid A_2} := \int dX\varrho_{\mathrm{ref}}(X)A_1^*(X)A_2(X),
\end{equation}
where the asterisk denotes complex conjugation.

We will be concerned with a continuum description of the particle density and
momentum current using the following definitions
\begin{equation}
  \rho(\vec r, t) = \frac1N\sum_i\delta\big(\vec r - \vec r_i(t)\big),\quad
  \vec j(\vec r, t) = \frac1N\sum_i\vec c_i\delta\big(\vec r - \vec r_i(t)\big).
\end{equation}
Note that the current $\vec j$ is defined with respect to the peculiar
velocities $\vec c_i$ \cite{chong+kim09,suzuki+hayakawa13}. We will
use the spatial Fourier transforms
$\rho_{\vec q}(t) = \mathsf{FT}[\rho](\vec q, t)$, and
$\vec j_{\vec q}(t) = \mathsf{FT}[\vec j](\vec q, t)$
\footnote{We use
 the convention
  $\mathsf{FT}[f](\vec q) = \int f(\vec r)e^{-i\vec q\cdot\vec
    r}d^3r$}.

\section{Granular Integration through Transients}
\label{sec:green-kubo}

The transient phase space distribution function evolves in time according to the Liouville equation
\begin{equation}
  \label{eq:22}
  \varrho(t) = \exp(-it\overline{\Lv}_+)\varrho_{\mathrm{ref}},
\end{equation}
Here $\overline{\Lv}_+$ is the adjoint of the full Liouville operator
$\Lv_+ = \Lv^+_{\mathrm{ref}} + \Lv_{\dot\gamma}$. This
together with Eq. (\ref{eq:8}) allows us \cite{fuchs+cates09} to rewrite the
sheared steady state phase space density $\varrho_{\mathrm{ss}}$ in
the following exact form,
\begin{equation}
  \label{eq:23}
  \varrho_{\mathrm{ss}} = \varrho_{\mathrm{ref}}
  + \int_0^{\infty}dt\exp(-it\overline{\Lv}_+)
  i\overline{\Delta\Lv}_+\varrho_{\mathrm{ref}},
\end{equation}
where
$\overline{\Delta\Lv}_+ = \Lv_{\dot\gamma} +
\overline{\Lv}^+_D(\Delta P_D)$,
and
$\overline{\Lv}^+_D(\Delta P_D) = -\Delta P_D\sum_j\partial^2_{\vec
  c_j}$.
The necessary reduction in driving power,
$\Delta P_D = P_D^{\dot\gamma} - P_D$, is to keep granular temperature
in the sheared stationary state fixed at the granular
temperature of the reference state, $T$.

We have argued in Sec.~\ref{sec:gener-green-kubo}, Eq.~(\ref{eq:9}),
that
$\varrho_{\mathrm{ref}}^{-1}i\Lv_{\dot\gamma}\varrho_{\mathrm{ref}}
= -\dot\gamma\sigma_{xy}^{\mathrm{el}}/T$
for a canonical approximation to $\varrho_{\mathrm{ref}}$. With the
same assumptions we find for the effect of the change in driving power
$ \varrho_{\mathrm{ref}}^{-1}i\overline{\Lv}^+_D(\Delta
P_D)\varrho_{\mathrm{ref}} = -3\Delta P_D\delta K/T$ where
\begin{equation}
  \label{eq:24}
  \delta K = 1 - \frac1N\sum_j\frac{\vec c_j^2}{3T}.
\end{equation}

Using Eq.~(\ref{eq:19}) to calculate the expectation value,
$\avr A_{\mathrm{ss}} := \int dX\varrho_{\mathrm{ss}}(X)A(X)$, of any
observable $A(X)$ in the sheared steady state, we find
\begin{equation}
  \label{eq:25}
    \avr A_{\mathrm{ss}} = \avr A
    - \frac{\dot\gamma}{T}\int_0^{\infty}dt\avr{\sigma_{xy}^{\mathrm{el}}\mid A(t)}
    - \frac{3\Delta P_{\mathrm{D}}}{T}\int_0^{\infty}dt\avr{\delta K\mid A(t)}.
\end{equation}
For any $A$ analytic in the velocities, the last term,
$\avr{\delta K\mid A(t)}$, will vanish due to the Gaussian property of
the velocity distribution. Formally, we establish a generalized
Green-Kubo relation which resembles the one for Brownian particles
\cite{fuchs+cates02,fuchs+cates09}
\begin{equation}
  \label{eq:26}
  \avr A_{\mathrm{ss}} = \avr A
  - \frac{\dot\gamma}{T}\int_0^{\infty}dt\avr{\sigma_{xy}^{\mathrm{el}}\mid A(t)}.
\end{equation}

Adapting an argument by \citet{chong+kim09}, we can show that for any
observable $A$, we have
\begin{equation}
  \label{eq:27}
  \avr{\sigma_{xy}^{\mathrm{el}}\mid\exp(it\Lv_+)A}
  = \avr{\sigma_{xy}^{\mathrm{el}}\Q\mid\exp(it\Q\Lv_+\Q)\Q A},
\end{equation}
where $\Q = 1-\P$ and
\begin{equation}
  \label{eq:28}
  \P = N\sum_{\vec q}\frac{1}{S_q}\ket{\rho_{\vec q}}\bra{\rho_{\vec q}}
  + \frac NT\sum_{\vec q}\ket{\vec j_{\vec q}}\bra{\vec j_{\vec q}}
\end{equation}
projects onto the hydrodynamic fields. Now, making a mode-coupling
approximation,
\begin{equation}
  \label{eq:29}
  \exp(it\Q\Lv_+\Q) \approx
  N^2\sum_{\vec k,\vec p}\ket{\rho_{\vec k(-t)}\rho_{\vec p(-t)}}
  \frac{\Phi_{\vec k(-t)}(t)\Phi_{\vec p(-t)}(t)}{S_{\vec k}S_{\vec p}}
  \bra{\rho_{\vec k}\rho_{\vec p}},
\end{equation}
we can express the Green-Kubo relation in terms of the transient correlator
$\Phi_{\vec q}(t)$, Eq.~(\ref{eq:11}),
\begin{equation}
  \label{eq:30}
  \avr A_{\mathrm{ss}} \approx \avr A
  - \frac{\dot\gamma}{2T}\sum_{\vec k}\int_0^{\infty}dt
  \mathcal V_{\vec k(-t)}^{\sigma}\mathcal W_{\vec k}^A\Phi_{\vec k(-t)}^2(t),
\end{equation}
where
\begin{equation}
  \label{eq:31}
  \mathcal W_{\vec k}^A = N\avr{\rho_{\vec k}\rho_{-\vec k}\mid A}/S_{\vec k}^2,
\end{equation}
\begin{equation}
  \label{eq:32}
  \mathcal V_{\vec k}^{\sigma}
  = N\avr{\sigma_{xy}^{\mathrm{el}}\mid \rho_{\vec k}\rho_{-\vec k}}
  = -T\frac{k_xk_y}{k}S'_k.
\end{equation}
where the prime denotes the derivative with respect to the wave vector.
The last equality has been derived from thermodynamic arguments in
Ref.~\cite{fuchs+cates02}. Here we present a kinetic derivation in
appendix~\ref{sec:density-stree}. Let us mention that
Eq.~(\ref{eq:30}) could be used to get the distorted structure under
shear choosing $A = |\rho_{\vec k}|^2$.

In particular, for $A = \sigma_{xy}$ we find
(cf.~appendix~\ref{sec:density-stree})
$\mathcal W_{\vec k}^{\sigma} = \frac{1 + \varepsilon}{2}\mathcal
V_{\vec k}^{\sigma}/S_k^2$,
\textit{i.e.}, for the macroscopic shear stress
\begin{equation}
  \label{eq:33}
  -\avr{\sigma_{xy}}_{\mathrm{ss}}/V = 
  \dot\gamma T\frac{1+\varepsilon}{4}
  \int\frac{d^3k}{(2\pi)^3}\int_0^{\infty}dt\frac{k_x^2k_yk_y(-t)}{kk(-t)}
  \times\frac{S'_{k(-t)}S'_k}{S_k^2}\Phi_{\vec k(-t)}^2(t).
\end{equation}
The above expression for the shear stress in terms of the transient
density correlation constitutes our first major result. The formula (\ref{eq:33}) for the granular shear stress and the
approximations required to derive it, closely corresponds to the
results and approximations for thermalized colloidal glass formers
\cite{fuchs+cates09}. While it is unknown how to systematically
improve Eq.~(\ref{eq:33}), it is based on physical intuition and has
been shown to be good for quiescent and shear driven dense Brownian
particle systems. Our result shares a number of features with the
\textsc{itt} expression for thermalized colloids. First, the
structural correlations encoded in the transient density correlator
determine the evolution of stresses under shear. Second, shearing
decorrelates density fluctuations by the advection to shorter
wavelengths where they decay more rapidly. The occurrence of
$\varepsilon$ quantifying the dissipation is specific to granular
media. It enters the mode-coupling vertex as found previously in
granular mode-coupling theory \cite{kranz+sperl13}. Besides
$\sigma_0$, which is neglected in dispersions, the role of temperature
differs in both approaches. While for thermalized Brownian particles,
the thermodynamic temperature is set by the heat bath and enters via
the initial canonical distribution, in the present approach to a
fluidized granular medium, the granular temperature $T$ is determined
from the power balance either in
\emph{Protocol H} or \emph{Protocol T}.

For thermalized colloids, two main approximations led to
Eq.~(\ref{eq:33}) with $\varepsilon=1$: (i) the focus on structural
relaxation, Eq.~(\ref{eq:29}), and (ii) the decoupling of higher order
correlations. For granular fluids, $\varepsilon < 1$, we had to make
mainly two additional approximations: (i) assume, Eq.~(\ref{eq:9}),
that the elastic Irving-Kirkwood shear stress is still the relevant
observable in the correlation function of the generalized Green-Kubo
integral, Eq.~(\ref{eq:10}), and (ii) assume the factorization between
spatial and velocity degrees of freedom in calculating the coupling
constants $\mathcal V_{\vec k}^{\sigma}$, Eq.~(\ref{eq:32}).

Note that the anisotropy of the sheared state is encoded in two ways
in Eqs.~(\ref{eq:33}): (i) by the anisotropic form of the coupling
constant, Eq.~(\ref{eq:32}), and (ii) by the anisotropy of the
advected wave vector $\vec k(-t)$.  Once we know the transient
correlator $\Phi_{\vec k}(t)$, this allows us to determine the shear
stress for a given shear rate $\dot\gamma$.  In terms of the
dimensionless time $t^* := \omega_ct$ and the dimensionless wave
number $k^* := kd$ we find a dimensionless stress
\begin{equation}
  \label{eq:34}
  -\avr{\sigma_{xy}}_{\mathrm{ss}}/NT
  = \Pe\frac{1+\varepsilon}{2\varphi}
  \int\frac{d^3k^*}{96\pi^2}\int_0^{\infty}dt^*\frac{k_x^{*2}k_y^*k_y^*(-t^*)}{k^*k^*(-t^*)}
  \times\frac{S'_{k^*(-t^*)}S'_{k^*}}{S_{k^*}^2}\Phi_{\vec k^*(-t^*)}^2(t^*)
\end{equation}
manifestly temperature independent.

\section{Mode-Coupling Theory for the Transient Correlator}
\label{sec:mode-coupling-theory}

The dynamics of the transient density correlator may be evaluated
self-consistently in the framework of mode-coupling theory. To do so,
we first isolate the conservation laws from the time evolution
(sec. \ref{sec:factoring-out-slow}) and subsequently formulate
equations of motion for the transient correlators $\Phi_{\vec q}(t)$
and $\vec H_{\vec q}(t)$ in terms of memory kernels
(Sec.~\ref{sec:equat-moti-trans}). These are approximated within
mode-coupling theory, resulting in closed equations for the transient
correlators (Sec.~\ref{sec:mode-coupl-appr}). The latter are solved
numerically, which can only be done within an isotropic approximation
(Sec.~\ref{sec:isotr-appr-1}).

\subsection{Separating the Slow Dynamics}
\label{sec:factoring-out-slow}

For the observables $A_{\vec q}\in\{\rho_{\vec q}, \vec j_{\vec q}\}$ we
define the propagator $\U(t)$ via
\begin{equation}
  \label{eq:35}
  A_{\vec q(t)}(t) = \exp(it\Lv_+)\exp(-it\Lv_{\dot\gamma_r})A_{\vec q}
  =: \U(t)A_{\vec q}
\end{equation}
where $\U(t)\ne\exp[it(\Lv_+ - \Lv_{\dot\gamma_r})]$ because $\Lv_+$ and
$\Lv_{\dot\gamma_r}$ do not commute. Following \citet{suzuki+hayakawa13} we
write
\begin{equation}
  \label{eq:36}
  \frac{d}{dt}\U(t) = \exp(it\Lv_+)i\tilde\Lv_+\exp(-it\Lv_{\dot\gamma_r})
\end{equation}
where $\tilde\Lv_+ = \Lv_+ - \Lv_{\dot\gamma_r}$. In order to separate
the slow dynamics, we define the projectors
\begin{equation}
  \label{eq:37}
    \P(t) :=
    N\sum_{\vec k}\ket{\rho_{\vec k(t)}}\bra{\rho_{\vec k(t)}}/S_{\vec k(t)}
    + N\sum_{\vec k}\ket{\vec j_{\vec k(t)}}\bra{\vec j_{\vec k(t)}}/T,
\end{equation}
and $\Q(t) = 1 - \P(t)$. Next, we can expand the time derivative of the
propagator as follows,
\begin{equation}
  \label{eq:38}
    \frac{d}{dt}\U(t) = \U_S(t)i\tilde\Lv_+\exp(-it\Lv_{\dot\gamma_r})
    + \Q(0)\hat\U(t, 0)\mathsf{R}(t) 
    + \int_0^td\tau\U_S(\tau)i\tilde\Lv_+\Q(\tau)\hat\U(t,\tau)\mathsf{R}(t),
\end{equation}
where
\begin{equation}
  \label{eq:39}
  \begin{aligned}
    \U_S(t) &:= \U(t)\exp(it\Lv_{\dot\gamma_r})\P(t)\\
    &= N\sum_{\vec k}\ket{\rho_{\vec k(t)}(t)}\bra{\rho_{\vec k(t)}}/S_{\vec k(t)}
    + N\sum_{\vec k}\ket{\vec j_{\vec k(t)}(t)}\bra{\vec j_{\vec k(t)}}/T
  \end{aligned}
\end{equation}
is the propagator of the slow modes,
\begin{equation}
  \label{eq:40}
  \hat{\U}(t,\tau) := \exp(-i\tau\Lv_{\dot\gamma_r}^{\dagger})
  \tilde\U(t,\tau)\exp(it\Lv_{\dot\gamma_r}),
\end{equation}
\begin{equation}
  \label{eq:41}
  \tilde\U(t,\tau) :=
  \exp_-\left[\int_{\tau}^tdt'\exp(it'\Lv_{\dot\gamma_r})\Q(t')
    i\tilde\Lv_+\exp(-it'\Lv_{\dot\gamma_r})\right],
\end{equation}
and $\exp_-$ is the time ordered exponential. Finally,
$\mathsf{R}(t) := \Q(t)i\tilde\Lv_+\exp(-it\Lv_{\dot\gamma_r})$, is
the fluctuating force operator. In the last term of Eq.~\eqref{eq:38},
the conventional small strain approximation \cite{fuchs+cates09}
$\exp(-it\Lv_{\dot\gamma_r}) \approx
\exp(-it\Lv_{\dot\gamma_r}^{\dagger})$
has been employed.  It reduces the consequences of shearing in the
equations of motion of $\Phi_{\vec q}(t)$ to the single effect of the
advection of wave vectors, as holds (without approximation) in the
formula (\ref{eq:34}) for the stress. Besides this approximation,
Eq.~\eqref{eq:38} is exact. Note that the second term in
Eq.~\eqref{eq:38} will never contribute due to the orthogonal
projector $\Q(0)$.

\subsection{Equation of Motion for the Transient Correlators}
\label{sec:equat-moti-trans}

The continuity equation for the density,
$i\tilde\Lv_+\rho_{\vec q(t)} = i\vec q(t)\cdot\vec j_{\vec q(t)}$,
implies
\begin{equation}
 \label{eq:42}
 \frac{d}{dt}\Phi_{\vec q}(t)
 = \vec q(t)\cdot\vec H_{\vec q}(t)/S_q
\end{equation}
where
\begin{equation}
  \label{eq:43}
  \vec H_{\vec q}(t) := iN\avr{\rho_{\vec q}\mid\vec j_{\vec q(t)}(t)}
\end{equation}
is the transient density-current correlator. Consider its time
derivative
\begin{equation}
  \label{eq:44}
  \frac{d}{dt}\vec H_{\vec q}(t)
  = iN\avr{\rho_{\vec q}\mid\frac{d}{dt}\mathsf{U}(t)\vec j_{\vec q}}
\end{equation}
which can be expanded with the help of Eq.~\eqref{eq:38}. For the first term we find
\begin{equation}
  \label{eq:45}
  \mathsf{U}_S(t)i\tilde\Lv_+\vec j_{\vec q(t)}
  = \vec{\Omega}^{\rho j}_{\vec q(t)}\rho_{\vec q(t)}(t)
  + \vec j_{\vec q(t)}(t)\cdot\mathsf{\Omega}_{\vec q(t)}
\end{equation}
where
\begin{subequations}
  \begin{align}
    \label{eq:46}
    \vec{\Omega}^{\rho j}_{\vec q}
    &:= N\avr{\rho_{\vec q}\mid i\tilde\Lv_+\vec j_{\vec q}}/S_q,\\
    \Omega^{\lambda\mu}_{\vec q}
    &:= N\avr{j_{\vec q}^{\lambda}\mid i\tilde\Lv_+j_{\vec q}^{\mu}}/T
  \end{align}
\end{subequations}
The third term yields for the fluctuating force $\vec R_{\vec q(t)} :=
\mathsf{R}(t)\vec j_{\vec q}$
\begin{equation}
  \label{eq:47}
  \mathsf{U}_S(\tau)i\tilde\Lv_+\Q(\tau)\hat{\mathsf{U}}(t,\tau)
  \vec R_{\vec q(t)}
  = -\vec L_{\vec q}(t,\tau)\rho_{\vec q(\tau)}(\tau)
  - \vec j_{\vec q(\tau)}(\tau)\cdot\mathsf{M}_{\vec q}(t,\tau)
\end{equation}
where
\begin{subequations}
  \begin{align}
    \label{eq:48}
    \vec L_{\vec q}(t,\tau) &:= -N\avr{\rho_{\vec q(\tau)}\mid
      i\tilde\Lv_+\Q(\tau)\hat{\mathsf{U}}(t,\tau)\vec R_{\vec q(t)}}/
    S_{\vec q(\tau)},\\
    M_{\vec q}^{\lambda\mu}(t,\tau) &:= -N\avr{j_{\vec q(\tau)}^{\lambda}\mid
      i\tilde\Lv_+\Q(\tau)\hat{\mathsf{U}}(t,\tau)R_{\vec q(t)}^{\mu}}/T
  \end{align}
\end{subequations}
are two memory kernels.

The elements of the frequency matrices are known from the literature:
$  \vec\Omega_{\vec q}^{\rho j} = i\vec qC^2_{\vec q}$ where
\begin{equation}
  \label{eq:49}
  C^2_{\vec q} = \frac{T}{S_q}\left[
    \frac{1+\varepsilon}{2} + \frac{1-\varepsilon}{2}S_q
  \right]
\end{equation}
is the (squared) speed of sound \cite{kranz+sperl13}, and
\begin{equation}
  \label{eq:50}
  \Omega_{\vec q} = \mu_{\vec q}\mathsf{1}+\mathsf{k}\quad
  \text{where \cite{kranz14}}\quad
  \mu_{\vec q} = \frac{1+\varepsilon}{3}\omega_c[1-j_0(qd)]
\end{equation}
and $j_0(x)$ is the zeroth order spherical Bessel function
\cite{gradshteyn+ryzhik00}.

The equations of motion for the correlators, $\Phi_{\vec q}(t)$ and
$\vec H_{\vec q}(t)$ can be written as
\begin{subequations}
  \label{eq:51}
\begin{equation}
  \label{eq:52}
  S_q\frac{d}{dt}\Phi_{\vec q}(t) = \vec q(t)\cdot\vec H_{\vec q}(t)
\end{equation}
and
\begin{equation}
  \label{eq:53}
  \begin{aligned}
    \frac{d}{dt}\vec H_{\vec q}(t)
    &+ \vec q(t)C^2_{\vec q(t)}S_q\Phi_{\vec q}(t)
    + \mu_{\vec q(t)}\vec H_{\vec q}(t) + \mathsf{k}\cdot\vec H_{\vec q}\\
    &+ \int_0^td\tau\vec L_{\vec q}(t, \tau)S_{\vec q(\tau)}\Phi_{\vec q}(\tau)
    + \int_0^td\tau\vec H_{\vec q}(\tau)\cdot\mathsf{M}_{\vec q}(t, \tau) = 0.
  \end{aligned}
\end{equation}
\end{subequations}

Were it not for the advected wave vector, Eqs.~(\ref{eq:51}) could
be combined into a single wave equation describing the propagation of
sound modes by taking the time derivative of
Eq.~(\ref{eq:52}). However because of the time-dependence of the wave
vector we obtain an additional term in
\begin{equation}
  \label{eq:54}
  S_q\ddot\Phi_{\vec q}(t) = \vec q\cdot\dot{\vec H}_{\vec q}(t)
  - \vec q(t)\cdot\mathsf{k}\cdot\vec H_{\vec q}(t),
\end{equation}
spoiling this simplification. It will nevertheless prove useful to
rewrite Eqs.~\eqref{eq:51} with help of the above relation
(Eq.\ref{eq:54}) in the form
\begin{equation}
  \label{eq:55}
  \begin{aligned}
    \ddot\Phi_{\vec q}(t) &+ \nu_{\vec q(t)}\dot\Phi_{\vec q}(t)
    + q^2(t)C^2_{\vec q(t)}\Phi_{\vec q}(t)
    + 2\vec q\cdot\mathsf{k}\cdot\vec H_{\vec q}(t)/S_q\\
    &+ \vec q(t)\cdot\int_0^td\tau\vec L_{\vec q}(t,\tau)
    \frac{S_{\vec q(\tau)}}{S_q}\Phi_{\vec q}(\tau)
    + \vec q(t)\cdot\int_0^td\tau\mathsf{M}_{\vec q}^{\mathsf t}(t,\tau)\cdot\vec
    H_{\vec q}(\tau)/S_q = 0
  \end{aligned}
\end{equation}
where \cite{kranz+sperl13}
\begin{equation}
  \label{eq:56}
  \nu_{\vec q} = \frac{1+\varepsilon}{3}\omega_c[1 + 3j''_0(qd)]
\end{equation}
is known as the Enskog term \footnote{For small wave numbers $q\to0$,
  and small densities, $n$, the Enskog term converges to
  $\nu_q\simeq9\zeta_Eq^2/5n$, where $\zeta_E$ is the Enskog
  expression for the bulk viscosity \cite{garzo+montanero02}} and the double primes denote the second
derivative. The first three terms in Eq.~(\ref{eq:55}) describe sound
waves with an advected wave vector $\vec q(t)$. The fourth term is
proportional to $\dot\gamma$ and vanishes in the limit of no
shear. All effects not local in time are contained in the memory
kernels $\vec L_{\vec q}$ and $\mathsf{M}_{\vec q}$. To make progress
we evaluate the memory kernels in terms of a mode-coupling
approximation.

\subsection{Mode-Coupling Approximation}
\label{sec:mode-coupl-appr}

We approximate $\hat\U(t,\tau)$ with a mode-coupling ansatz
\cite{fuchs+cates09},
\begin{equation}
  \label{eq:57}
  \begin{aligned}
    \hat\U(t,\tau) &\approx\P_2(\tau)\hat\U(t,\tau)\P_2(t)\\
    &\approx N^2\sum_{\vec k,\vec p} \ket{\rho_{\vec
        k(\tau)}\rho_{\vec p(\tau)}} \frac{\Phi_{\vec
        k(\tau)}(t-\tau)\Phi_{\vec p(\tau)}(t-\tau)} {S_{\vec
        k(t)}S_{\vec p(t)}} \bra{\rho_{\vec k(t)}\rho_{\vec p(t)}},
  \end{aligned}
\end{equation}
where
\begin{equation}
  \label{eq:58}
  \P_2(t) := N^2\sum_{\vec k,\vec p}
  \ket{\rho_{\vec k(t)}\rho_{\vec p(t)}}
  \bra{\rho_{\vec k(t)}\rho_{\vec p(t)}}/S_{\vec k(t)}S_{\vec p(t)}.
\end{equation}
This immediately yields $\vec L_{\vec q}(t,\tau)\approx\vec 0$ as the
left vertex
$\propto\avr{\rho_{\vec q(\tau)}\mid i\tilde\Lv_+\Q(\tau)\rho_{\vec
    k(\tau)}\rho_{\vec p(\tau)}} = 0$ due to parity.
For the second memory kernel we find
\begin{equation}
  \label{eq:59}
  M_{\vec q}^{\lambda\mu}(t,\tau) \approx
  \frac NT\sum_{\vec k,\vec p}\frac{S_{\vec k(\tau)}}{S_{\vec k(t)}}
  \mathcal V_{\vec q\vec k\vec p}^{\lambda}(\tau)
  \mathcal W_{\vec q\vec k\vec p}^{\mu}(t)
  \Phi_{\vec k(\tau)}(t-\tau)\Phi_{\vec p(\tau)}(t-\tau)
\end{equation}
Note that that the memory kernel, Eq.~(\ref{eq:59}), is defined in
terms of the \emph{transient} correlators
$\Phi_{\vec k(\tau)}(t-\tau) \equiv \Phi_{\vec k(\tau)}(0, t-\tau)$.
Although the appearance of the relative time $t-\tau$ suggests an
assumption of time translation invariance, no such assumption is made,
and, in fact, it would be incorrect during the transient onset of
shear.  The vertices,
\begin{subequations}
  \begin{align}
    \label{eq:60}
    \mathcal V_{\vec q\vec k\vec p}^{\lambda}(t)
    &= N\avr{j_{\vec q(t)}^{\lambda}\mid i\tilde\Lv_+\Q(t)\rho_{\vec
        k(t)}\rho_{\vec p(t)}}/S_{\vec k(t)},\\
    \mathcal W_{\vec q\vec k\vec p}^{\lambda}(t)
    &= N\avr{\rho_{\vec k(t)}\rho_{\vec
        p(t)}\mid\Q(t)i\tilde\Lv_+j_{\vec q(t)}^{\lambda}}/S_{\vec p(t)}
  \end{align}
\end{subequations}
are, again, known from the literature \cite{kranz+sperl13}. They
depend on time only parametrically because of the advected wave vector
[$\mathcal{V}_{\vec q\vec k\vec p}(t)\equiv\mathcal{V}_{\vec q(t)\vec
  k(t)\vec p(t)}$].
Therefore, the calculation relating them to the quiescent static
structure factor of hard spheres is unaffected,
\begin{subequations}
  \label{eq:61}
  \begin{align}
    \label{eq:62}
    i\mathcal V_{\vec q\vec k\vec p}^{\lambda} &= \frac TNS_{\vec p}
    [k^{\lambda}nc_{\vec k} + p^{\lambda}nc_{\vec p}]\delta_{\vec q,\vec k+\vec p},\\
    i\mathcal W_{\vec q\vec k\vec p}^{\lambda}
    &= \frac{1 + \varepsilon}{2}\frac TNS_{\vec k}
    [k^{\lambda}nc_{\vec k} + p^{\lambda}nc_{\vec p}]\delta_{\vec q,\vec k+\vec p}.
  \end{align}
\end{subequations}
This closes the equations of motion as soon as we know the static
structure factor $S_{\vec q}$.

The solution of Eqs.~(\ref{eq:51}) together with Eq.~(\ref{eq:59}) and
the explicit expressions for the vertices, Eqs. (\ref{eq:61}), provides a
non-trivial prediction of the transient correlator $\Phi_{\vec
  q}(t)$. In particular, it captures the glass transition of the
quiescent granular fluid for vanishing shear \cite{kranz+sperl10}.  It
is by no means obvious that a granular fluid, where the dissipative
interactions violate detailed balance, is described by equations of
motion that share the same structure as the equations of motion for a
fluid comprised of elastic, thermalized particles \cite{chong+kim09},
and are straight forwardly related to the equations for Brownian
suspensions \cite{fuchs+cates09}. Note, however, that the speed of
sound, $C_{\vec q}$, the viscosity term
$\mu_{\vec q}$, and, most importantly, the memory kernel $\mathsf{M}_{\vec
  q}(t,\tau)$ all depend on the coefficient of restitution
$\varepsilon$. As discussed in Ref.~\cite{kranz+frahsa17a}, the
prescribed shear rate always melts the glass but the interplay of the
structural relaxation rate and the shear rate still determines a large
part of the rheology through the generalized Green-Kubo relation
(\ref{eq:33}). The self-consistent equations for the transient correlators---valid within the approximations
outlined above for a granular fluid at high densities and arbitrary
shear rates---are our second major result. Written in terms of
dimensionless time, $t^* := \omega_ct$, and wave vector, $\vec k^* :=
\vec
kd$, these equations are manifestly temperature independent, testament
to the fact that changes in granular temperature cannot change the
qualitative behavior of hard sphere fluids.

\subsection{The Isotropic Approximation}
\label{sec:isotr-appr-1}

Applying shear in a prescribed direction breaks the isotropy of the
fluid and this is reflected in the anisotropy of
Eqs.~(\ref{eq:51}). Note that we require isotropy in terms of the wave
vector $\vec k$ at time $t$, not in terms of the initial wave vector
$\vec k(-t)$. The wave vector advection is then taken into account in
a directionally averaged fashion as well
\cite{fuchs+cates03,chong+kim09},
$\frac{1}{4\pi}\int d\uvec k\vec k^2(-t) =: {\bar k}^2(-t)$,
\textit{i.e.}, $\bar k(-t) = k\sqrt{1 + (\dot\gamma t)^2/3}$. With an
isotropic $\Phi_{\bar k(-t)}(t)$, the Green-Kubo relation
(\ref{eq:33}) will only depend on the isotropic part of the
vertex-product
\begin{equation}
  \label{eq:63}
  \frac{1}{4\pi}\int d\uvec k
  \mathcal V^{\sigma}_{\vec k(-t)}\mathcal V^{\sigma}_{\vec k}
  = \frac{T^2k^2}{15\sqrt{1 + (\dot\gamma t)^2/3}}S'_{\bar k(-t)}S'_k,
\end{equation}
\textit{i.e.}, with an isotropic transient correlator $\Phi_k(t)$, the
generalized Green-Kubo relation for a sheared granular fluid reads
\footnote{For a dimensionless version, see Eq.~(4) of Ref.~\cite{kranz+frahsa17a}}
\begin{equation}
  \label{eq:64}
  \sigma = \sigma_0 + \frac{1 + \varepsilon}{2}\dot\gamma T
  \int_0^{\infty}\frac{dt}{\sqrt{1 + (\dot\gamma t)^2/3}}
  \int_0^{\infty}\frac{dkk^4}{60\pi^2}\times
  \frac{S'_{\bar k(-t)}S'_k}{S_k^2}
  \Phi^2_{\bar k(-t)}(t).
\end{equation}

Averaging Eq.~(\ref{eq:52}) over the directions $\uvec q(t)$ we have
\begin{equation}
  \label{eq:65}
  S_q\dot\Phi_q(t) = q(t)\uvec q(t)\cdot\vec H_q(t)
\end{equation}
which shows that the density correlator only couples to the
longitudinal part, $\uvec q(t)\cdot\vec H_q(t)$, of the
density-current correlator.
Thereby $\vec H_{\vec q}(t)$ is eliminated as an independent
dynamical quantity in Eqs.~(\ref{eq:51}) and
allows recovery of the non-linear wave equation
(cf.~appendix~\ref{sec:direct-aver-eq}),
\begin{equation}
  \label{eq:66}
  \ddot\Phi_q(t) + \nu_{q(t)}\dot\Phi_q(t) + q^2(t)C_{q(t)}^2\Phi_q(t)
  + q^2(t)C_{q(t)}^2\int_0^td\tau m_{\vec q}(t,\tau)\dot\Phi_q(\tau) = 0,
\end{equation}
a single scalar integro-differential equation replacing the four
equations (\ref{eq:51}). Here the damping term $\nu_q$,
Eq.~(\ref{eq:56}), and the speed of sound, $C_q$, have become
isotropic and the dimensionless longitudinal part of the memory
kernel,
$m_{\vec q}(t, \tau) := \vec q(t)\vec q(t):\mathsf{M}_{\vec
  q}(t,\tau)/q^2(t)C_{q(t)}^2$,
is given below by Eq.~(\ref{eq:67}). Had we, instead, assumed that the
directional distribution of the initial wave vectors, $\vec q$, were
uniform and let it evolve into a non-uniform distribution of wave
vectors $\vec q(t)$ at time $t$, the damping term $\nu_q$ would be
supplemented by an additional, wave-number independent contribution
\cite{suzuki+hayakawa13}.

In isotropic approximation we have to leading order
$[\uvec q(t)\cdot\vec k(t)]/q(t) \simeq \uvec q\cdot\vec k/q$ and find
\begin{multline}
  \label{eq:67}
  m_{\vec q}(t,\tau) = A_{\bar q(t)}(\varepsilon)\frac{S_{\bar q(t)}}{nq^2}
  \int\frac{d^3k}{(2\pi)^3}S_{\bar k(\tau)}S_{\bar p(\tau)}\\
  \times[(\uvec q\cdot\vec k)nc_{\bar k(t)} + (\uvec q\cdot\vec
  p)nc_{\bar p(t)}]
  [(\uvec q\cdot\vec k)nc_{\bar k(\tau)} + (\uvec q\cdot\vec
  p)nc_{\bar p(\tau)}]\\
  \times\Phi_{\bar k(\tau)}(t-\tau)\Phi_{\bar p(\tau)}(t-\tau).
\end{multline}
For vanishing shear, $\dot\gamma\to0$, the memory kernel reduces to
the one calculated in Ref.~\cite{kranz+sperl13} and
\begin{equation}
  \label{eq:68}
  A_q^{-1}(\varepsilon) = 1 + \frac{1-\varepsilon}{1+\varepsilon}S_q.
\end{equation}

\begin{figure}[t]
  \centering
  \includegraphics{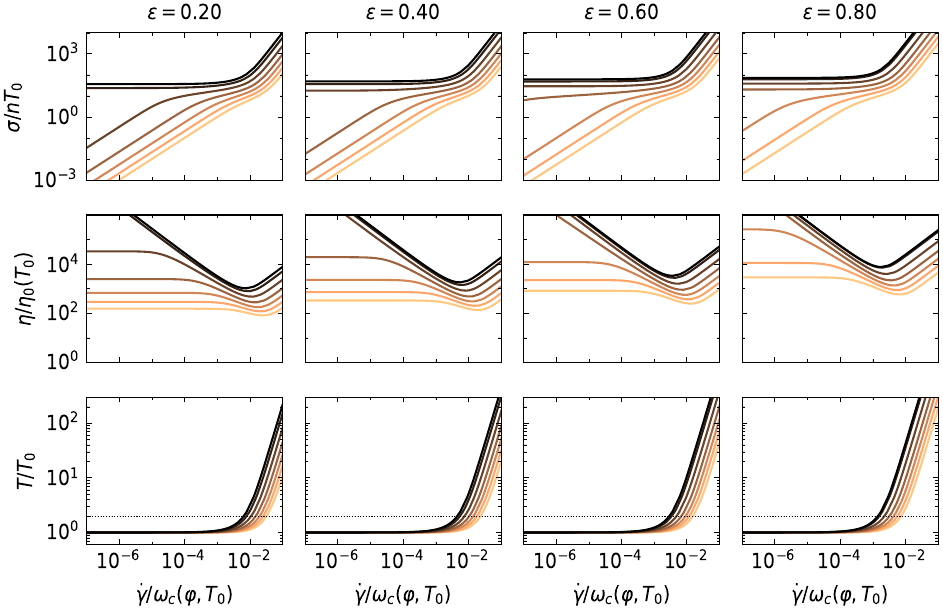}
  \caption{\textbf{Protocol H} (Shear heating increases the granular
    temperature). For several values of the coefficient of restitution
    $\varepsilon$ (columns, as indicated) and several packing fractions (color
    coded) from $\varphi=0.48$ (yellow, bottom) to $\varphi=0.60$ (black,
    top): (First row) Flow curves, shear stress $\sigma$ as a function of
    shear rate $\dot\gamma$, normalized by the density $n$ and the granular
    temperature at zero shear, $T_0$. (Second row) Viscosity $\eta$ relative
    to the Boltzmann viscosity $\eta_0(T_0)$, Eq.~(\ref{eq:69}) as a function
    of shear rate $\dot\gamma$. (Third row) Stationary granular temperature
    $T$, relative to the initial granular temperature, $T_0$, at zero shear as
    a function of shear rate $\dot\gamma$. The dotted lines indicate
    $T/T_0\equiv2$. The shear rate is normalized by the collision frequency
    $\omega_c(\varphi, T_0)$ of the unsheared system (at granular temperature
    $T_0$) with the prescribed volume fraction.}
  \label{fig:intrinsic}
\end{figure}

\subsection{Numerical Solutions}
\label{sec:numerical-solutions}

The granular mode-coupling equations (\ref{eq:66}, \ref{eq:67}) can in
general only be solved numerically. The same applies to the
generalized Green-Kubo relation (\ref{eq:64}). To this end we need the
static structure factor and the sound damping constant in the
reference state. For the structure factor we resort to the
Percus-Yevick expression \cite{percus+yevick58} for elastic hard
spheres to be consistent with the discussion of the granular glass
transition \cite{kranz+sperl10,kranz+sperl13}. One could just as well
use structure factors from numerical simulations of the reference
state, given they are available at sufficient quality. The dependence
on the coefficient of restitution manifests itself primarily in the
height of the first peak. For the sound damping constant we used the
Enskog expression of \citet{garzo+montanero02}. It is not a crucial
ingredient but more precise values could easily be incorporated. For
more technical details on the numerical procedure see
appendix~\ref{sec:numerics}.

\section{Results}
\label{sec:results}

To obtain quantitative predictions, we numerically solved
Eqs.~(\ref{eq:66}, \ref{eq:67}, \ref{eq:64}) (see
Sec.~\ref{sec:numerical-solutions} and
appendix~\ref{sec:numerics}). Some of the results are presented in
Ref.~\cite{kranz+frahsa17a}. Here we will focus on a broader range of
parameters and provide additional comparison with established kinetic
theories.

\subsection{Rheology \& Flow Curves}
\label{sec:rheology--flow}

\subsubsection{Protocol H}
\label{sec:protocol-h}

\begin{figure}[t]
  \centering
  \includegraphics[width=6.4in]{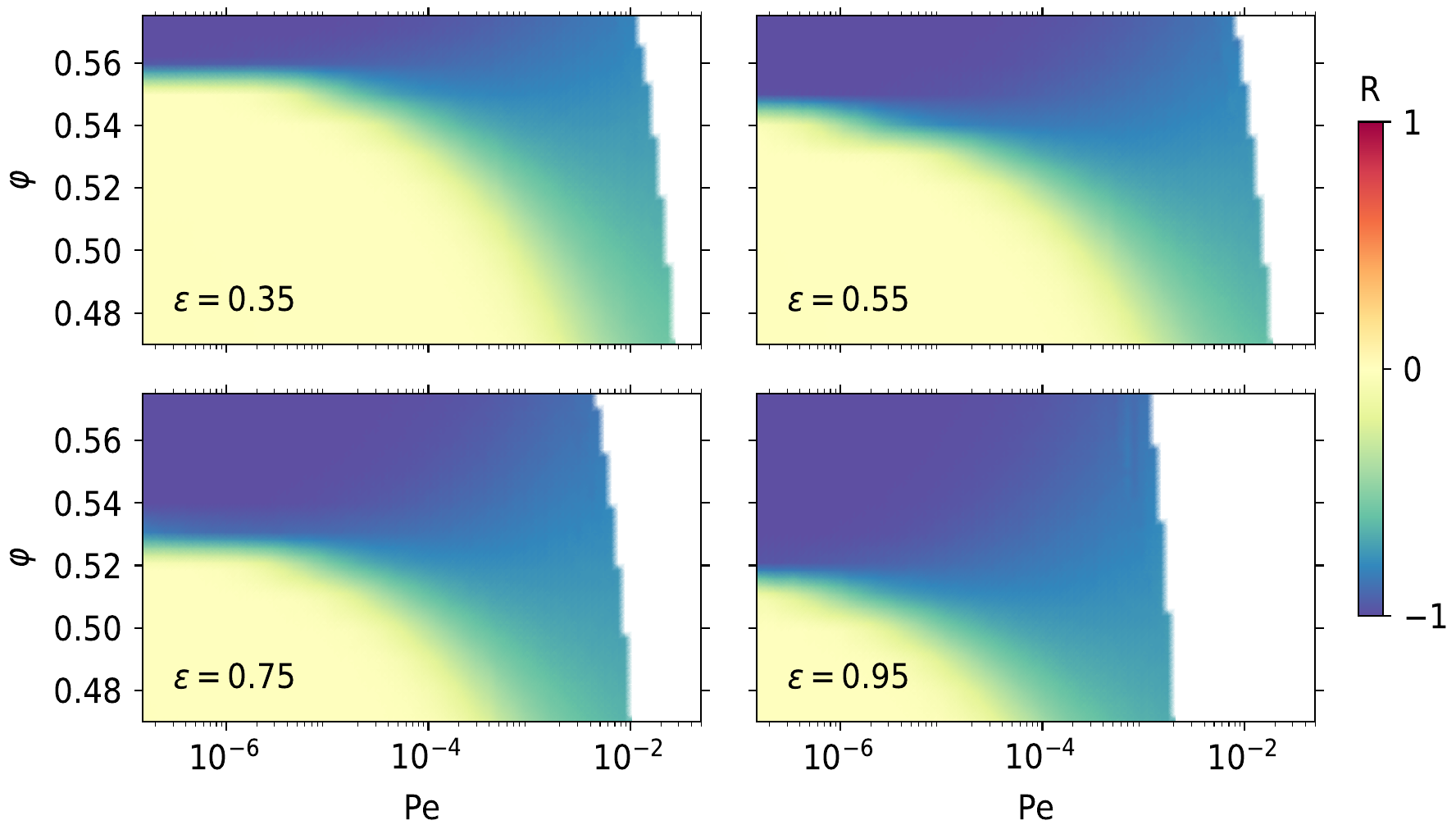}
  \caption{\textbf{Protocol T} ($T\equiv T_0$) or \textbf{Protocol H} ($T >
    T_0$): Dynamic state diagrams in the plane spanned by P\'eclet number,
    $\Pe$, and packing fraction, $\varphi$, for several values of the
    coefficient of restitution, $\varepsilon$, as indicated. The effective
    exponent $R$ quantifying the shear rate dependence of the viscosity,
    $\eta(\Pe) \sim \Pe^R$, is color coded on the same scale as in
    Fig.~\ref{fig:regimes}. The inaccessible regime $\Pe > \Pe_{\infty}$ is
    left blank. The jagged boundary is due to discretization.}
  \label{fig:regimespe}
\end{figure}

\begin{figure}[t]
  \centering
  \includegraphics{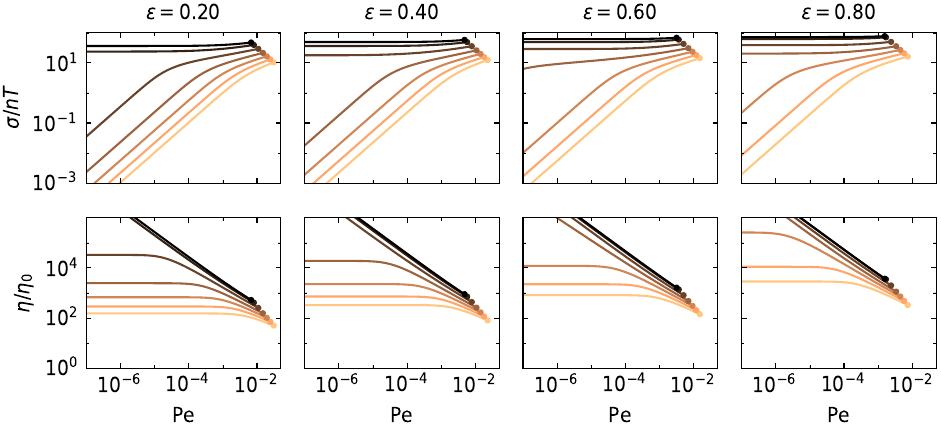}
  \caption{\textbf{Protocol T} ($T\equiv T_0$) or \textbf{Protocol H}
    ($T > T_0$): Flow curves, shear stress $\sigma$ as a function of
    P\'eclet number $\Pe$ (first row), and viscosity $\eta$ relative
    to the Boltzmann viscosity $\eta\equiv\eta_0(T)$,
    Eq.~(\ref{eq:69}), (second row) for several values of the
    coefficient of restitution $\varepsilon$ (columns, as indicated)
    and several packing fractions (color coded) from $\varphi=0.48$
    (yellow, bottom) to $\varphi=0.60$ (black, top). The filled
    circles mark the maximum P\'eclet number $\Pe_{\infty}$,
    \textit{i.e.}, the end of the flow curves for the respective
    densities. }
  \label{fig:intrinsicpe}
\end{figure}

\begin{figure}[t]
  \centering
  \includegraphics{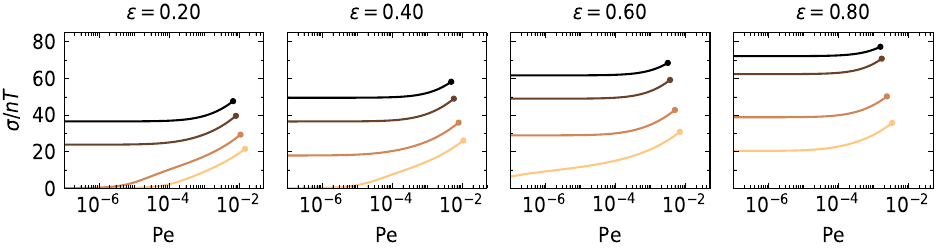}
  \caption{\textbf{Protocol T} ($T\equiv T_0$) or \textbf{Protocol H} ($T >
    T_0$): Flow curves, $\sigma(\Pe)$, for several values of the coefficient
    of restitution $\varepsilon$ (as indicated) and several packing fractions
    (color coded) from $\varphi=0.54$ (yellow, bottom) to $\varphi=0.60$
    (black, top) close to and above the glass transition density. Note the
    linear scale for the shear stress.}
  \label{fig:sigmape}
\end{figure}

The phenomenology of \emph{Protocol H} has been discussed in
Ref.~\cite{kranz+frahsa17a}. Comparing the dynamic state diagrams for
different inelasticities, $\varepsilon$, in Fig.~\ref{fig:regimes}
confirms the broad layout of the rheological regimes. The critical
density for the granular glass transition, $\varphi_c(\varepsilon)$,
which increases with increasingly dissipative particles determines the
boundaries of the Newtonian regime observed at low densities and small
shear rates. At the highest shear rates, the power balance,
Eq.~(\ref{eq:4}), is dominated by shear heating and shear thickening
is observed in the Bagnold regime. As the granular fluid is more
susceptible to shear heating the more elastic the particles are, the
onset of the Bagnold regime moves to lower shear rates for larger
values of $\varepsilon$.

Flow curves, $\sigma(\dot\gamma)$, corresponding to \emph{Protocol H}
are shown in Fig.~\ref{fig:intrinsic} (see also Figs.~2 and 3 in
Ref.~\cite{kranz+frahsa17a}). Note that the precise value of the
coefficient of restitution may have a comparatively large influence on
the flow behavior. For the same flow conditions, \textit{i.e.},
packing fraction, $\varphi$, and shear rate, $\dot\gamma$, rather
elastic particles may place the granular fluid in the flat part of the
flow curve, requiring rather large stresses, $\sigma/nT_0 \sim 10$,
while for more inelastic particles the flow would still be in the
Newtonian regime, requiring only negligible shear stress,
$\sigma/nT_0 \ll 1$.  This is also reflected in the viscosity
(Fig.~\ref{fig:intrinsic}) which may vary over many orders of
magnitude for a fixed packing fraction, $\varphi$, depending on both
shear rate and coefficient of restitution. A natural scale to compare
viscosities to is the Boltzmann viscosity,
\begin{equation}
\label{eq:69}
  \eta_0 := \eta(\varphi\to0, \varepsilon=1) = \frac{5}{16d^2}\sqrt{T/\pi},
\end{equation}
of an elastic hard sphere gas at vanishing density
\cite{chapman+cowling70}. Fig. \ref{fig:intrinsic} shows that viscosities far
exceeding this value are predicted at the considered
densities. Considering the stationary granular temperature $T$,
resulting from the power balance, Eq.~(\ref{eq:4}), we observe that,
as expected, significant shear rates are needed to make shear heating
relevant and increase the stationary granular temperature, $T$, with
respect to the granular temperature $T_0$ of the unsheared fluid
(Fig.~\ref{fig:intrinsic}). In Ref.~\cite{kranz+frahsa17a} we defined
the critical P\'eclet number, $\Pe^*$, for the onset of shear
thickening, as the point where $T=2T_0$. Comparing the temperature
curves in Fig.~\ref{fig:intrinsic} with the viscosity curves,
$\eta(\dot\gamma)$, confirms the utility of this definition.

If we normalize the shear stress, $\sigma/nT$, with the stationary granular
temperature, $T$, instead of the initial granular temperature, $T_0$, and
express the shear rate, $\dot\gamma$, in terms of the P\'eclet number, $\Pe$,
we arrive at a description of the rheology in terms of intrinsic quantities
that make the temperature independence of the (inelastic) hard sphere fluid
manifest (Figs.~\ref{fig:regimespe},\ref{fig:intrinsicpe}). In this form, the
predictions for \emph{Protocol H} are identical to the ones for \emph{Protocol
  T} as they only differ in the temperature control
(cf.~Sec.~\ref{sec:flow-problem}).

\subsubsection{Protocol T}
\label{sec:protocol-t}

Employing \emph{Protocol T} makes it straightforward to obtain the intrinsic
quantities, $\sigma/nT$, and $\Pe = \dot\gamma/\omega_c(\varphi, T)$, as the
granular temperature is held constant, $T\equiv T_0$, throughout. In terms of
the packing fraction, $\varphi$, and the P\'eclet number, $\Pe$, the dynamic
state diagram is shown in Fig.~\ref{fig:regimespe} for different
inelasticities $\varepsilon$. We readily observe that the shear thickening
regime vanishes altogether compared to the diagrams for \emph{Protocol H}
(Fig.~\ref{fig:regimes}). The apparent shear thickening observed in
\emph{Protocol H} is due to shear heating only, which is absent in
\emph{Protocol T}. As the P\'eclet number is restricted to be smaller than the
maximal value, $\Pe \leq \Pe_{\infty}$ \cite{kranz+frahsa17a}, the dynamic
state diagram includes unreachable regions of large P\'eclet number,
$\Pe > \Pe_{\infty}(\varphi, \varepsilon)$ that cannot be realized in a
granular fluid. The Bagnold regime of \emph{Protocol H}
(cf.~Fig.~\ref{fig:regimes}) shrinks to the line
$\Pe = \Pe_{\infty}(\varphi, \varepsilon)$ for \emph{Protocol T}. At low
P\'eclet number or shear rate, where shear heating is negligible, the dynamic
state diagrams become indistinguishable between \emph{Protocols H} and
\emph{T} (cf.~Figs.~\ref{fig:regimes} and \ref{fig:regimespe}). Shear thinning
behavior, which is caused by the slow relaxation in the fluid
\cite{kranz+frahsa17a,kranz+sperl10} is independent of temperature control and
is therefore also observed in \emph{Protocol T}.

The flow curves, $\sigma(\Pe)$, for \emph{Protocol T} shown in
Fig.~\ref{fig:intrinsicpe} (see also Fig.~2 in
Ref.~\cite{kranz+frahsa17a}) also reflect the finite admissible range
of P\'eclet numbers as they end at $\Pe = \Pe_{\infty}$. The viscosity
curves (Fig.~\ref{fig:intrinsicpe}) confirm that no shear thickening
is observed in \emph{Protocol T} and the Newtonian regime at low
densities and small shear rates is complemented by a shear thinning
regime only. For packing fractions,
$\varphi > \varphi_c(\varepsilon)$, above the glass transition, the
variation of the shear stress, $\sigma$, with P\'eclet number, $\Pe$,
is remarkably small throughout the whole range of admissible P\'eclet
numbers (Fig.~\ref{fig:sigmape}). Once the yield stress,
$\sigma_y = \sigma(\Pe\to0)$, is exceeded, no more than roughly a
doubling of the shear stress will be needed to achieve arbitrary shear
rates.

\subsection{Transport Coefficients and Yield Stress}
\label{sec:transp-coeff-}

\begin{figure}[t]
  \centering
  \includegraphics{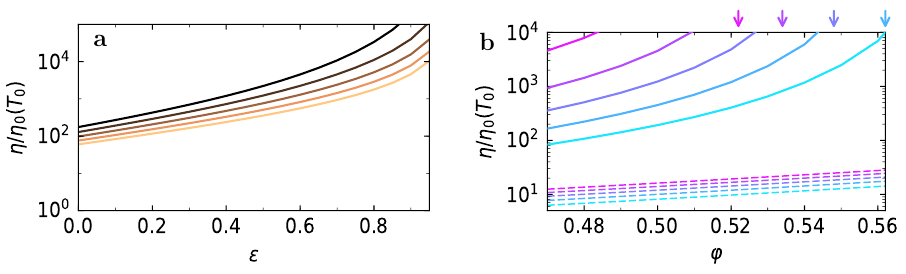}
  \caption{Viscosity $\eta$, normalized by the Boltzmann viscosity
    $\eta_0$, Eq.~(\ref{eq:69}), at small shear rate,
    $\dot\gamma\to0$. (a) As a function of the coefficient of
    restitution $\varepsilon$ for a number of packing fractions from
    $\varphi=0.47$ (yellow, bottom) to $\varphi=0.51$ (black,
    top). (b) As a function of packing fraction, $\varphi$, for a
    number of values of the coefficient of restitution, $\varepsilon$,
    from $\varepsilon=0.1$ (cyan, bottom) to $\varepsilon=0.9$
    (magenta, top). Solid lines denote \textsc{gitt} predictions while
    dashed lines indicate the Enskog low density expansion by
    \citet{garzo+montanero02}. The arrows indicate the glass
    transition density, $\varphi_c(\varepsilon)$.}
  \label{fig:eta}
\end{figure}

Irrespective of the possibility to handle arbitrary shear rates,
\textsc{gitt} (and more generally \textsc{itt}) can also be used to
calculate the viscosity, $\eta$, in the linear response regime,
$\dot\gamma\to 0$. Recall that in this limit \emph{Protocol T} and
\emph{H} become indistinguishable as there is no shear
heating. Employed in this way, \textsc{gitt} extends the low density
Enskog predictions \cite{garzo+montanero02}, to higher densities
(Fig.~\ref{fig:eta}). In particular it captures the strong increase
and eventual divergence of the viscosity,
$\eta(\varphi)\propto
(\varphi_c(\varepsilon)-\varphi)^{-\gamma(\varepsilon)}$ with
$\gamma(\varepsilon)\sim2.4$ as the glass transition is approached
\cite{kranz+frahsa17a,kranz+sperl13}. This divergence can, of course,
not be recovered by the low density, Enskog predictions. The
variations in the glass transition, $\varphi_c(\varepsilon)$, with the
coefficient of restitution, $\varepsilon$, leads to a correspondingly
large variation of the viscosity, $\eta(\varepsilon)$, which far
exceeds the variations predicted by Enskog theory.

\begin{figure}[t]
  \centering
  \includegraphics{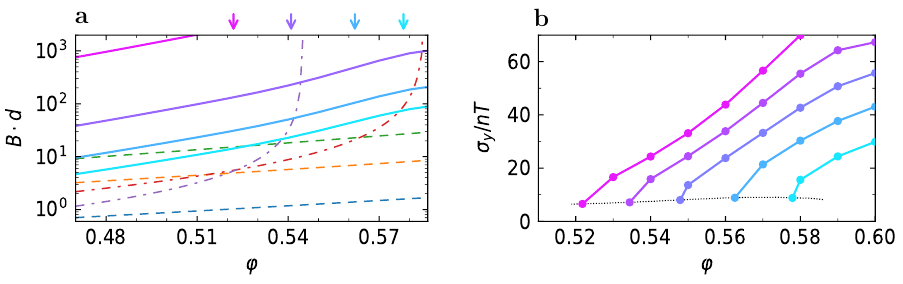}
  \caption{(a) Bagnold coefficient $B$ as a function of packing
    fraction $\varphi$. Solid lines are \textsc{gitt} predictions for
    a number of values of the coefficient of restitution,
    $\varepsilon$, from $\varepsilon = 0.1$ (cyan, bottom) to
    $\varepsilon=0.9$ (magenta, top). The dashed lines denote Enskog
    predictions from \citet{savage+jeffrey81} (blue),
    \citet{kumaran06} (orange), and \citet{mitarai+nakanishi05}
    (green). The last two are for $\varepsilon=0.9$. The arrows
    indicate the glass transition density,
    $\varphi_c(\varepsilon)$. The dot-dashed lines are the predictions
    by Jenkins \textit{et al.} \cite{jenkins+berzi10,jenkins07}
    $B_{J1}$ (red) and $B_{J2}$ (violet) for $\varepsilon=0.6$
    (cf.~appendix~\ref{sec:jenk-const-equat}). (b) Yield stress,
    $\sigma_y$, as a function of volume fraction, $\varphi$, for a
    number of values of the coefficient of restitution, $\varepsilon$,
    from $\varepsilon=0.9$ (magenta, left) to $\varepsilon=0.1$ (cyan,
    right). Points are calculated and lines are a guide to the
    eye. The dotted horizontal line indicates the critical yield
    stress $\sigma_y^c$ at the granular glass transition.}
  \label{fig:bagnold}
\end{figure}

The large shear rate behavior, $\dot\gamma\to\infty$, or
$\Pe\to\Pe_{\infty}$, respectively, is characterized by the Bagnold
coefficient $B := \sigma/{\dot\gamma}^2$, Fig.~\ref{fig:bagnold}a. As
shear heating is more effective for more elastic particles (in
\emph{Protocol H}), the Bagnold coefficient increases with
$\varepsilon$. As the glass is already shear-molten in the Bagnold
regime, unlike the viscosity $\eta$, the Bagnold coefficient does not
diverge at the glass transition density,
$\varphi_c(\varepsilon)$. However, the increasing sluggishness of the
fluid at high densities is also reflected in the Bagnold coefficient
which increases rapidly with density. Such a quick rise is naturally
not recovered by the existing Enskog predictions. The theories by
\citet{mitarai+nakanishi05}, and \citet{kumaran06} both come close to
the results from \textsc{gitt} for lower densities but deviate by
orders of magnitude close to the glass transition density. The simple
prediction by \citet{savage+jeffrey81},
$ B_{\mathrm{SJ}}d = 32\varphi\chi(\varphi)/35\pi$, where
$\chi(\varphi)$ is the value of the pair correlation function at
contact is independent of the coefficient of restitution but also
yields the right order of magnitude at low densities.

While the
Bagnold coefficient of \citet{savage+jeffrey81} is expressed in terms
of the pair correlation function at contact of elastic hard spheres,
namely, the Carnahan-Starling approximation \cite{hansen+mcdonald90}
that is very good up to $\varphi\approx0.4$, \citet{jenkins07} later
used an empirical, high-density pair-correlation function that
diverges at $\varphi_{\mathrm{rcp}}\approx 0.64$ to predict a Bagnold coefficient
$B_{J1}$, Eq.~(\ref{eq:102}), that is formally close to Savage and
Jeffrey's expression, but depends on the coefficient of restitution
$\varepsilon$ and diverges at $\varphi_{\mathrm{rcp}}$.
Note that, as discussed in Ref.~\cite{kranz+frahsa17a},
we underestimate the granular glass transition by roughly $5.5\%$,
corresponding to $\Delta\varphi\sim 0.055$ for $\varepsilon=0.6$. To
compare Jenkins' results to ours, we consequently display his result
for $B_{J1}(\varphi - 0.055)$ in Fig.~\ref{fig:bagnold} at a shifted
density, so that it now diverges at $\varphi=0.585$.
In a later paper, \Citet{jenkins+berzi10} put forward the idea of a
lower shear jamming density at $\varphi\approx0.6$ for
$\varepsilon\approx0.7$, implying a divergence of the Bagnold
coefficient $B_{J2}$ at $\varphi\approx0.6$,
Eq.~(\ref{eq:106}). Again shifting density for comparison with our results,
we show $B_{J2}(\varepsilon=0.6, \varphi - 0.055)$ in
Fig.~\ref{fig:bagnold}a.

In contrast, the granular mode-coupling theory here applies to a
driven granular fluid and predicts a granular glass transition and the
emergence of a yield stress at a critical density
$\varphi_c(\varepsilon)<\varphi_{\mathrm{rcp}}$.  The discussion of
the relevant time scales \cite{kranz+frahsa17a} and the dynamic state
diagrams, cf. Fig.~\ref{fig:regimes} shows that the shear rate in the
Bagnold regime, $\dot\gamma_{\infty}$, exceeds the structural
relaxation rate, $\tau_{\alpha}^{-1}$. Therefore, the slow relaxation,
$\tau_{\alpha}\to\infty$ indicating the granular glass transition, is
irrelevant for the Bagnold regime and the Bagnold coefficient must be
smooth across the granular glass transition. This is, indeed, the case
for the Bagnold coefficient $B$ calculated by the \textsc{gitt}
formalism (Fig.~\ref{fig:bagnold}a).

Above the glass transition density, $\varphi_c(\varepsilon)$, a finite
(dynamic) yield stress, $\sigma_y$, has to be exceeded to keep the
system flowing and to prevent it from freezing into an amorphous
glass. The critical yield stress, $\sigma_y^c(\varepsilon)$
(cf.~Fig.~3 in Ref.~\cite{kranz+frahsa17a}) right at the glass
transition is on the order of $6$--$9nT$. For larger densities,
$\varphi > \varphi_c(\varepsilon)$, the yield stress quickly rises
with density (Fig.~\ref{fig:bagnold}b) as expected \footnote{To
  calculate $\sigma_y$, a Herschel-Bulkley law,
  $\sigma(\Pe) = \sigma_y + \alpha\Pe^n$ \cite{herschel+bukley26}, is
  fitted to the numerically determined flow curves for
  $\Pe \leq 10^{-4}$.}.

\section{Conclusion}
\label{sec:conclusion}

In summary, we have shown how to derive generalized Green-Kubo
relations, Eq.~(\ref{eq:30}), for sheared frictionless inelastic hard
sphere fluids at finite shear rates and densities around the
(granular) glass transition density. In particular, we have shown that
an \textsc{itt} formalism can be derived for a granular fluid which is
generically out of euilibrium, even in the absence of shear.  Here, we
focused on the generalized Green-Kubo relation for the shear stress
$\sigma$, Eq.~(\ref{eq:33}), that we found to be formally identical to
the one for thermalized colloidal suspensions, Eq.~(37) in
Ref.~\cite{fuchs+cates09}, but for the prefactor
$(1+\varepsilon)/2$. The starting point was the exact
relation~(\ref{eq:8}), expressing the phase space distribution of the
sheared stationary state, $\varrho_{\mathrm{ss}}$, in terms of the
distribution of the quiescent but out-of-equilibrium reference state,
$\varrho_{\mathrm{ref}}$. The essential approximations were: (i) the
assumption that the microscopic shear stress captures the change of
$\varrho_{\mathrm{ref}}$ due to shear, Eq.~(\ref{eq:9}), and (ii) to
replace the decay of stress correlations under the Green-Kubo integral
by the slow relaxation of density fluctuations encoded in the
transient correlator $\Phi_{\vec q}(t)$, Eq.~(\ref{eq:12}). The
coupling constants, Eq.~(\ref{eq:32}), capture the essential shear
induced anisotropy making it sufficient to consider an isotropic
transient correlator $\Phi_q(t)$, Eq.~(\ref{eq:64}).  The derivation
showed that the generalized Green-Kubo relation~(\ref{eq:12}) is
neither restricted to small shear rates, $\dot\gamma\to0$, nor to
quasi-elastic particles, $\varepsilon\to1$. To ensure a clear
separation of time scales between the slow relaxation of density
fluctuations and all other relaxation modes in the fluid, the density
needs to be not too far below the granular glass transition density
$\varphi_c(\varepsilon)$ \cite{kranz+sperl10}. On the upper end, it
should be sufficiently far away from the random close packing density,
$\varphi_{\mathrm{rcp}}\approx0.64$, as we do not take the jamming
transition into account. Coveniently, details of the driving mechanism
do no longer feature explicitly in the generalized Green-Kubo
relation~(\ref{eq:12}).

For a homogeneously cooling granular gas, a Green-Kubo relation akin
to Eq.~(\ref{eq:10}) has been developed by
\citet{goldhirsch+vannoije00} following Ronis' \cite{ronis79}
perturbative approach but they have kept the stress auto-correlation
function as an irreducible function. \Citet{kumaran09c} has derived a
Green-Kubo relation similar to our result, Eq.~(\ref{eq:12}), for a
dilute quasi-elastic sheared granular gas in the Bagnold regime. His
result is perturbative in the small parameter
$\sqrt{1-\varepsilon}$. While it incorporates wave vector advection
and employs a factorization of the time evolution operator similar to
our mode-coupling approximation, it uses the linearized granular
hydrodynamic equations valid at low density and small wave number,
$kd\ll1$.

To capture the slow glassy relaxation in the quiescent fluid that is a
result of strongly correlated dynamics and sensitive to local particle
configurations, $kd\gtrsim1$
\cite{debenedetti+stillinger01,kranz+sperl13}, we had to go beyond
linearized hydrodynamic equations. Instead, we derived a set of
non-linear, generalized hydrodynamic equations for the transient
correlator, Eqs.~(\ref{eq:51},\ref{eq:59}), explicitly taking into
account the shear advection of wave vectors and the loss of detailed
balance due to the inelastic collisions. The latter is manifest in the
modified speed of sound, Eq.~(\ref{eq:49}), and the two unequal
coupling constants in the memory kernel,
$\mathcal V^{\lambda}_{\vec q\vec k\vec p},\mathcal W^{\lambda}_{\vec
  q\vec k\vec p}$, Eqs.~(\ref{eq:61}), see also
Refs.~\cite{kranz+sperl10,kranz+sperl13}. To this end, we kept the
exact inelastic binary collision operator, Eq.~(\ref{eq:17}), but
approximated the non-equilibrium phase space distribution
$\varrho_{\mathrm{ref}}$ of the reference fluid as essentially the
canonical distribution of an elastic hard sphere fluid
(cf.~Sec.~\ref{sec:flow-problem}). Let us once more stress that here
the essential approximation is the factorization between spatial and
velocity degrees of freedom. A quantitative understanding of these
correlations is currently lacking and numerical results are limited to
the low density, two-dimensional data of
\citet{pagonabarraga+trizac01}. A fully quantitative treatment of the
coupling between relative positions and velocities is therefore
currently out of reach. However, in order to assess the validity of
our approximations, let us sketch the qualitative influence of these
correlations on the \textsc{gitt} formalism. As the correlations are
short ranged \cite{pagonabarraga+trizac01}, we expect them to
primarily affect the collision integrals used to calculate the
vertices. In effect, the mean collisional force is reduced through the
depletion of head-on collisions. This has two counteracting effects:
(i) It reduces the shear stress at a given density, but (ii) it shifts
the granular glass transition to higher densities. In conclusion, we
expect the shear stress at a given distance from the granular glass
transition to be only weakly dependent on the coupling between
positions and velocities. Non-Gaussian corrections to the velocity
distribution as well as structure factors that characterize the actual
granular fluid could be incorporated in extensions of our work.

A notable feature of the \textsc{gitt} approach presented here is that
it associates the emergence of a yield stress with the dynamic
granular glass transition. The corresponding glass transition
densities, $0.52\lesssim\hat\varphi_c(\varepsilon)\lesssim0.6$
\cite{kranz+sperl10,kranz+frahsa17a}, are lower than the eventual
jamming density at $\varphi_{\mathrm{rcp}}\approx0.64$. In essence, we
predict the emergence of a yield stress (characterizing a quiescent
amorphous solid) without lasting contacts between the particles. Shear
thinning is then associated with the shear rate $\dot\gamma$ exceeding
the structural relaxation rate of the granular fluid
\cite{kranz+frahsa17a}. The latter is a hallmark of glassy behavior
but is at variance with alternative approaches to granular rheology
that attribute the yielding transition at densities below
$\varphi_{\mathrm{rcp}}$ with the shear induced emergence of lasting
contacts
\cite{reddy+kumaran07,kumaran09a,kumaran09b,kumaran14,jenkins+berzi10},
\textit{i.e.}, a divergence of the pair correlation function at
contact, $\chi$. Resolving which of the mechanisms controls the
physics, requires, we believe, additional experiments and simulations.

The increase of the shear stress at high densities is driven by the
slow structural relaxation of the system around the glass
transition. The dependence of the glass transition on the coefficient
of restitution yields a strong sensitivity of the transport
coefficients and even the qualitative rheological behavior on the
inelasticity of the particles. This sensitivity, together with the
strong increase (and eventual divergence) of the small shear
viscosity, $\eta(\dot\gamma\to0)$, shows that an extrapolation of the
Enskog predictions from the gaseous state of vanishing density,
$\varphi\ll1$, to significant densities, $\varphi\sim\mathcal O(1)$,
cannot work as it completely neglects the dramatic slowing down of
structural relaxation at high densities. For the Bagnold coefficient,
$B$, we have argued that it is only defined for finite shear rates,
$\dot\gamma_{\infty} = \omega_c\Pe_{\infty}$. Except for the elastic
limit, $\varepsilon\to1$, where $\Pe_{\infty} \ll 1$, this makes the
Bagnold coefficient inaccessible to linear response theories but
places it well within the reach of \textsc{gitt}.

The hard-sphere property that the temperature, $T$, only enters as a timescale
via the collision frequency, $\omega_c\sim\sqrt T$, and does not control the
physics, is retained by \textsc{gitt}. In terms of the P\'eclet number, $\Pe$,
and the dimensionless shear stress $\sigma/nT$, the theory is
manifestly temperature independent. The choice of temperature control in an
experiment, however, has a profound influence on the phenomenology
observed. If shear heating is not compensated (\emph{Protocol H}), the work
expended on heating the system will manifest as an apparent shear thickening
(Fig.~\ref{fig:regimes}). However, the viscosity $\eta\sim\sqrt T$ trivially
rises with temperature and if we take this into account
(Fig.~\ref{fig:regimespe}), no shear thickening remains. Dialing down the
random driving force we can keep the temperature constant (\emph{Protocol T}),
However, this only works below a maximal shear rate, $\dot\gamma \leq
\omega_c\Pe_{\infty}$. For higher shear rates even switching off the random
driving completely cannot compensate shear heating.

We have focused here on the derivation of \textsc{gitt} and only
discussed the most fundamental predictions of the theory. The
experience from the rheology of colloidal suspensions shows that the
\textsc{itt} formalism can contribute to the description of a much
broader range of phenomena. We are planning to explore these in future
work. Straightforward extensions include a more realistic, velocity
dependent coefficient of restitution \cite{schwager+poeschel08}. We do
not expect such a refined model to qualitatively change our
results. On the other hand, the rheology of granaular fluids is
strongly effected
\cite{heussinger+grob14,heussinger+grob16,wyart+cates14} by frictional
interactions of the grains, which could be taken into account by
tangential restitution and Coulomb friction as in
Refs.~\cite{herbst+huthmann00,herbst+cafiero05}.  Another extension
are larger size disparities, which may entail new phenomena
\cite{imhof+dhont95} and are left for subsequent studies. Our
formalism could also be extended to treat non-spherical particles
\cite{aspelmeier+huthmann01}, albeit at the expense of a further
increase in technical complexity. A necessary first step would be to
promote the understanding of the orientational
\cite{schilling+scheidsteger97,theenhaus+schilling01,chong+goetze02,elizondo+rico14}
glass transition to the granular realm. Tackling this formidable task
is far beyond the scope of the present study and shall be left to
future work. In the present work, we have neglected drag forces and
hydrodynamic interactions induced by a possible interstitial fluid and
effectively consider particles in vacuum, leaving a discussion of
two-phase flow effects to future work.

\begin{acknowledgments}
  We acknowledge crucial insight from discussions with Hisao Hayakawa,
  Koshiro Suzuki, Thomas Voigtmann, and Claus Heussinger. We thank
  three anonymous referees for outstanding reports that greatly
  improved the paper. We are grateful for the detailed feedback on
  this revision provided by Olivier Coquand. We thank the \textsc{dfg}
  for partial funding through FOR\ 1394 and KR\ 4867/2.
\end{acknowledgments}

\appendix

\section{The Sllod Equations of Motion}
\label{sec:sllod-equat-moti}

The Sllod equations form the basis of the work presented here. As they
posses some non-intuitive features and have been a matter of debate in
the literature
\cite{evans+morriss07,edwards+dressler01,edwards+baig06,daivis+todd06},
we will sketch their derivation. To make contact with the literature,
let's assume for the moment that particles interact via smooth pair
potentials $V(r_{ij})$. Then the Hamiltonian of the system is
generically given as
$H(\vec r_1,\ldots,\vec r_N,\vec v_1,\ldots,\vec v_N) = \sum_iv_i^2/2
+ \sum_{i<j}V(r_{ij})$.

To single out the imposed flow profile $\vec u(\vec r) = \vec
r\cdot\mathsf{k}$, we make the non-canonical transformation $\vec
v_i\to\vec c_i = \vec v_i - \vec r_i\cdot\mathsf{k}$. In the new
coordinates the Hamiltonian
\begin{equation}
  \label{eq:70}
  H_{\mathrm{nc}}(\vec r_1,\ldots,\vec r_N,\vec c_1,\ldots,\vec c_N)
  = \sum_i\left(\frac12c_i^2 + \vec r_i\vec c_i:\mathsf{k}\right)
  + \tilde V(\vec r_1,\ldots,\vec r_N)
\end{equation}
assumes the form of a Dolls-Hamiltonian, albeit with a a modified
effective potential
$\tilde V(\vec r_1,\ldots,\vec r_N) := \sum_{i<j}V(r_{ij}) +
\frac12\sum_i(\vec r_i\cdot\mathsf{k})^2$ \cite{edwards+dressler01}.
The canonical Hamilton equations no longer hold for the non-canonical
Hamiltonian, Eq.(\ref{eq:70}), expressed in non-canonical
coordinates. Instead, the same transformation has to be applied to the
canonical Poisson brackets,
$\{\cdot,\cdot\} \to \{\cdot,\cdot\}_{\mathrm{nc}}$ \cite[Eq.~(22)
in][]{edwards+dressler01}. The resulting equations of motion are the
Sllod equations
\begin{align}
  \label{eq:71}
  \dot{\vec r}_i &= \{\vec r_i, H_{\mathrm{nc}}\}_{\mathrm{nc}}
                   = \vec c_i + \vec r_i\cdot\mathsf{k}\\
  \label{eq:72}
  \dot{\vec c}_i &= \{\vec c_i, H_{\mathrm{nc}}\}_{\mathrm{nc}}
                   = \vec F_i - \vec c_i\cdot\mathsf{k}
                   - \vec r_i\cdot\mathsf{k}\cdot\mathsf{k}
\end{align}
where $\vec F_i := -\sum_{j\ne i}\nabla V(r_{ij})$ is the total force
on particle $i$.

Note that the net forces are not independent of the flow
profile. Indeed one can show that the time averaged force
$\overline{\vec F}(\vec r) := \overline{\sum_{i\in O}\vec F_i}$ acting
on a fluid element $O$ at position $\vec r$ is fully determined by the
velocity gradient tensor,
$\overline{\vec F}(\vec r) = \vec r\cdot\mathsf{k}\cdot\mathsf{k}$
\cite{edwards+baig06}. Therefore any fluctuations
$\delta\vec u(\vec r) := \vec u(\vec r) - \vec r\cdot\mathsf{k} =
\sum_{i\in O}\vec c_i$ follow the simple equation of motion
\begin{equation}
  \label{eq:73}
  \delta\dot{\vec u}(\vec r, t) = -\delta\vec u(\vec r, t)\cdot\mathsf{k}
  + \delta\vec F(\vec r, t)
\end{equation}
where $\delta\vec F(\vec r, t) = \vec F(\vec r, t) - \overline{\vec
  F}(\vec r)$ is the fluctuating force. The first term on the right
hand side of Eq.~(\ref{eq:73}) is problematic. Depending on the
velocity gradient tensor $\mathsf{k}$ it may lead to unbounded growth
of a spontaneous fluctuation $\delta\vec u(\vec r, 0)$. In simulations
this is avoided by taking the whole simulation volume as the fluid
element $O$ and using deterministic boundary forces (\textit{e.g.},
Lees-Edwards boundary conditions) with vanishing fluctuations
$\delta\vec F = \vec 0$. Together with a perfect initial flow field,
$\delta\vec u(\vec r, t=0) = \vec 0$, this keeps the fluctuations in
the flow field at exactly zero for all times as the total momentum of
the system
\begin{equation}
  \label{eq:74}
  \sum_i\vec c_i(0) = \vec 0
\end{equation}
is conserved.

For interacting particles, one expects viscous heating, \textit{i.e.},
the transfer of energy from the flow field $\vec r\cdot\mathsf{k}$ to
the internal energy
$E_{\mathrm{int}} := \sum_ic_i^2/2 + \sum_{i<j}V(r_{ij}) \ne
H_{\mathrm{nc}}$. Indeed, one finds
\begin{equation}
  \label{eq:75}
  \frac{dE_{\mathrm{int}}}{dt} = -\sigma:\mathsf{k} - \dot W_{\mathrm{exp}}
\end{equation}
where
$W_{\mathrm{exp}} = \sum_i\vec r_i\vec c_i:\mathsf{k}\cdot\mathsf{k}$
is the work performed by the fluid due to an expansive flow field
\cite{edwards+baig06}.  As it is usually desired to keep the
temperature---related to the internal energy---of the fluid constant,
a thermostat has to be employed \cite{evans+morriss07}.  In the
present work we are solely concerned with linear shear flows where
$\nabla\vec u\cdot\nabla\vec u \equiv 0$. simplifying both
Eqs.~(\ref{eq:72}) and (\ref{eq:75}).

\section{The Density-Stress-Overlap for Hard Spheres}
\label{sec:density-stree}

To make contact with established results, it is instructive to recall
the origin of the momentum current tensor $\mathsf{P}$ in the
continuity equation for momentum
\begin{equation}
  \label{eq:76}
  \partial_t\vec j(\vec r, t) \equiv i\Lv^+_{\mathrm{ref}}\vec j(\vec r, t)
  = -\nabla\cdot\mathsf{P}(\vec r, t).
\end{equation}
In reciprocal space we therefore have \cite{wajnryb+altenberger95}
\begin{equation}
  \label{eq:77}
  -i\vec q\cdot\mathsf{P}(\vec q)
  = i\vec q\cdot\frac{\uvec q\uvec q}{iq}i\Lv_0j_{\vec q}^L
  + i\vec q\cdot\sum_{j<k}\frac{\uvec r_{jk}}{i\vec q\cdot\uvec r_{jk}}
  i\mathcal T^+_{jk}\vec j_{\vec q},
\end{equation}
where $j_{\vec q}^L = \uvec q\cdot\vec j_{\vec q}$ is the longitudinal
current. In the absence of macroscopic flows ($\vec u\equiv\vec0$)
such that $\sigma_{\alpha\beta} \equiv -P_{\alpha\beta}$ and in the
homogeneous limit $\vec q\to0$ this indeed recovers the
Irving-Kirkwood expressions $\sigma_{\alpha\beta}^{\mathrm{kin}}(\vec q\to0) =
\sum_jv_j^{\alpha}v_j^{\beta}$, and
\begin{align}
  \label{eq:78}
  \sigma_{\alpha\beta}^{\mathrm{int}}(\vec q\to0)
  &= -\frac{1+\varepsilon}{2}\sum_{j<k}
    \frac{\hat r_{jk}^{\alpha}\hat r_{jk}^{\beta}}{i\vec q\cdot\uvec r_{jk}}
    (\uvec r_{jk}\cdot\vec v_{jk})^2\theta(-\uvec r_{jk}\cdot\vec v_{jk})
    \delta(r_{jk} - d)\left(e^{i\vec q\cdot\vec r_k} - e^{i\vec q\cdot\vec r_j}\right)\\
  &= \frac{1+\varepsilon}{2}\sum_{j<k}
    \frac{i\vec q\cdot\vec r_{jk}}{i\vec q\cdot\uvec r_{jk}}
    \hat r_{jk}^{\alpha}\hat r_{jk}^{\beta}
    (\uvec r_{jk}\cdot\vec v_{jk})^2\theta(-\uvec r_{jk}\cdot\vec v_{jk})
    \delta(r_{jk} - d)\\
  \label{eq:79}
  &= \frac{1+\varepsilon}{2}d\sum_{j<k}\hat r_{jk}^{\alpha}\hat r_{jk}^{\beta}
    (\uvec r_{jk}\cdot\vec v_{jk})^2\theta(-\uvec r_{jk}\cdot\vec v_{jk})
    \delta(r_{jk} - d),
\end{align}
independent of $\uvec q$ \cite{dufty12,baskaran+dufty08}. The vertex
$\mathsf{A} := -\avr{\rho_{\vec k}\rho_{-\vec k}\mid\Q\mathsf{P}(\vec
  q = \vec 0)} = \uvec k\uvec kA_{\parallel} + (\mathsf{1} - \uvec
k\uvec k)A_{\perp}$ may be split into a longitudinal and a transverse
part
\begin{equation}
  \label{eq:80}
  A_{\parallel} = \uvec k\uvec k:\mathsf{A},\quad\text{and}\quad
  A_{\perp} = \uvec k^{\perp}\uvec k^{\perp}:\mathsf{A},\quad\text{and}\quad
\end{equation}
where $\uvec k^{\perp}\perp\uvec k$. Then the off-diagonal component
is given by $A_{xy} = \uvec k_x\uvec k_y(A_{\parallel} - A_{\perp})$.

In order to calculate the expectation values $A_{\parallel}$ and
$A_{\perp}$, we start with the stress tensor at finite $\vec q$ and
only perform the limit $q\to0$ in the end. For $A_{\parallel}$
momentum conservation enforces $\vec q\parallel\vec k$ such that
\begin{equation}
  \label{eq:81}
  A_{\parallel} = -\lim_{q\to0}
  \avr{\rho_{\vec k}\rho_{q\uvec k-\vec k}\mid\Q\uvec k\uvec k:\mathsf{P}(q\uvec k)}
\end{equation}
where
\begin{equation}
  \label{eq:82}
  -\uvec k\uvec k:\mathsf{P}(q\uvec k)
  = \frac{1}{iq}i\Lv_0j_{q\uvec k}^L
  + \sum_{j<k}\frac{\uvec k\cdot\vec r_{jk}}{iq\uvec k\cdot\vec r_{jk}}
  i\mathcal T^+_{jk}j_{q\uvec k}^L = \frac{1}{iq}i\Lv^+_{\mathrm{ref}}j_{q\uvec k}^L,
\end{equation}
\textit{i.e.},
\begin{equation}
  \label{eq:83}
  A_{\parallel} = \lim_{q\to0}\frac{1}{iq}
  \avr{\rho_{\vec k}\rho_{q\uvec k-\vec k}\mid\Q i\Lv^+_{\mathrm{ref}}j_{q\uvec k}^L}.
\end{equation}
With the same line of argument we arrive at
\begin{equation}
  \label{eq:84}
  A_{\perp} = -\lim_{q\to0}
  \avr{\rho_{\vec k}\rho_{q\uvec k^{\perp}-\vec k}\mid\Q\uvec k^{\perp}\uvec k^{\perp}:
    \mathsf{P}(q\uvec k^{\perp})}
  = -\lim_{q\to0}\frac{1}{iq}
  \avr{\rho_{\vec k}\rho_{q\uvec k^{\perp}-\vec k}\mid\Q i\Lv^+_{\mathrm{ref}}j_{q\uvec k^{\perp}}^L}.
\end{equation}
From the literature \cite{kranz+sperl13} we know
\begin{equation}
  \label{eq:85}
  \avr{\rho_{\vec k}\rho_{\vec q-\vec k}\mid\Q i\Lv^+_{\mathrm{ref}}j_{\vec q}^L}
  = \frac{1+\varepsilon}{2}iT\left[
    (\uvec q\cdot\vec k)S_{|\vec q - \vec k|}
    + \uvec q\cdot(\vec q - \vec k)S_k - qS^{(3)}(\vec k, \vec q - \vec k)/S_q
  \right],
\end{equation}
where $S^{(3)}(\vec k, \vec p)$ is the triplet static structure factor
\cite{rosenfeld+levesque90}. The magnitude of the off-diagonal term
can therefore be calculated as
\begin{equation}
  \label{eq:86}
  \begin{aligned}
    A_{\parallel} - A_{\perp} &= \frac{1+\varepsilon}{2}T
    \lim_{q\to0}\left[
      k\frac{S_{k-q} - S_k}{q}
      + S_k - \frac{S^{(3)}(\vec k, q\uvec k - \vec k)}{S_q} - S_k
      + \frac{S^{(3)}(\vec k, q\uvec k^{\perp} - \vec k)}{S_q}
    \right]\\ &
    = -\frac{1+\varepsilon}{2}TkS'_k
  \end{aligned}
\end{equation}
employing the geometric relations between $\vec q$ and $\vec k$ and
using that $S(q\to0)\ne0$. The latter is true for both elastic and
dissipative hard sphere fluids in the stationary state but fails in
hyperuniform configurations \cite{torquato18}.  For elastic spheres
($\varepsilon=1$) this constitutes a kinetic derivation of the
stress-density overlap that has previously been derived from
thermodynamic arguments \cite{fuchs+cates09}.

\section{The Case of Viscous Dissipation}
\label{sec:case-visc-diss}

Besides modeling granular particles as hard spheres with a coefficient
of restitution $\varepsilon$, an often used and equally valid approach
is to model them as soft spheres with a finite ranged, elastic
repulsive force $\vec F_i^{\mathrm{el}}(r_{ij})$
\cite{poeschel+schwager05}. In this case, dissipation is modeled by a
viscous force
\begin{equation}
  \label{eq:87}
  \vec F_i^{\mathrm{visc}} = -\zeta\sum_{j\ne i}\Theta(d - r_{ij})
  \uvec r_{ij}(\vec v_{ij}\cdot\uvec r_{ij})
\end{equation}
acting while the particles are in contact and parameterized by the
viscosity $\zeta$, a material property of the particles. Note that
$\vec F_i^{\mathrm{visc}}$ is linear in the relative velocity.

A Liouville operator corresponding to $\vec F_i^{\mathrm{visc}}$ is
easily constructed,
$i\Lv^{\mathrm{visc}} = \sum_i\vec
F_i^{\mathrm{visc}}\cdot\partial_{\vec v_i}$. If this Liouville
operator is used in Eqs.~(\ref{eq:83}) and (\ref{eq:84}) above, it can
be seen that the viscous contributions vanish due to parity. With
this, and in contrast to our expression~(\ref{eq:12}), the generalized
Green-Kubo relation for soft granular particles with viscous
dissipation does not contain any explicit reference to the dissipative
interactions. In fact, it is identical to the relation for purely
elastic particles. For the same reason, also the equation of motion
for the transient correlator $\Phi_{\vec k}(t)$ including its memory
kernel in mode-coupling approximation does not contain any term that
stems from the viscous dissipation. The only change in the
constitutive relation between purely elastic soft particles and
dissipative soft particles would then be contained in the change of
static structure, $S_k$, and is expected to be very weak.

To remedy this unsatisfactory state, \citet{suzuki+hayakawa14} amended
the mode-coupling approximation of the stress auto-correlation
function, Eq.~(\ref{eq:10}), by a second projection on density-current
pairs $\rho_{\vec k}\vec j_{-\vec k}$. These have a finite overlap
with the viscous stress $i\Lv^{\mathrm{visc}}\vec j_{\vec
  k}$. However, this forces the introduction of a second, independent
transient correlator,
$\overline{\vec H}_{\vec k}(t) := iN\avr{\vec j_{\vec k}\mid\rho_{\vec
    k(t)}(t)}$ in our notation. Even in the isotropic approximation,
$\overline{\vec H}_{\vec k}(t)$ cannot be reduced to the transient
density correlator due to the broken time-reversal symmetry. The
second correlator, coming with its own equation of motion coupled to
the equation of motion of $\Phi_{\vec k}(t)$, substantially increases
the complexity of the problem. Although \citet{suzuki+hayakawa14}
have shown that it is still manageable, it is not completely clear
that to single out density-current pairs is physically
consistent. Density-density pairs are unique in that the (transient)
correlator is expected to become exceedingly slow compared to
microscopic time scales close to the glass transition. Currents,
however, do not freeze at the glass transition and current
correlations are expected to decay rapidly. At this point, it is not
obvious if there are other decay channels in addition to
$\overline{\vec H}_{\vec k}(t)$ that would be equally important and
would have to be included.

\section{The Directional Average of Eq.~(\ref{eq:54})}
\label{sec:direct-aver-eq}

The following derivation is standard in the \textsc{itt} literature
\cite{fuchs+cates09,chong+kim09,suzuki+hayakawa13} but we include it
here for the sake of completeness. The vector valued density-current
correlator
$\vec H_{\vec q}(t) = \uvec q(t)H^{\parallel}_{\vec q}(t) + \uvec
q^{\perp}(t)H^{\perp}_{\vec q}(t)$
has a longitudinal and a transverse component. Here
$\uvec q^{\perp}(t)\perp\uvec q(t)$ is an \emph{a priori} arbitrary
unit vector perpendicular to $\uvec q(t)$. From Eq.~(\ref{eq:65}) we
have $H_q^{\parallel}(t) = S_q\dot\Phi_q(t)/q(t)$; furthermore $\vec L_{\vec q}=0$ in mode-coupling approximation, so that
substitution of Eq.~(\ref{eq:53}) into Eq.~(\ref{eq:54}) yields
\begin{multline}
  \label{eq:88}
  S_q\ddot\Phi_{\vec q}(t) + q^2(t)C^2_{\vec q(t)}S_q\Phi_{\vec q}(t)
  + q(t)\nu_{\vec q(t)}H_{\vec q}^{\parallel}(t)
  + \vec q(t)\cdot
  \int_0^td\tau\vec H_{\vec q}(\tau)\cdot\mathsf{M}_{\vec q}(t, \tau)\\
  + 2\vec q(t)\cdot\mathsf{k}\cdot\vec H_{\vec q}(t) = 0.
\end{multline}
Once we also split the memory kernel into longitudinal and transverse
components
\begin{equation}
  \label{eq:89}
  \mathsf{M}_{\vec q}(t,\tau)
  = \uvec q(t)\uvec q(t)m_{\vec q}^{\parallel}(t, \tau)
  + [\mathsf{1} - \uvec q(t)\uvec q(t)]m_{\vec q}^{\perp}(t, \tau)
\end{equation}
we have
\begin{equation}
  \label{eq:90}
  \vec q(t)\cdot[\vec H_{\vec q}(\tau)\cdot\mathsf{M}_{\vec q}(t,\tau)]
  = q(t)H_{\vec q}^{\parallel}(\tau)m_{\vec q}^{\parallel}(t, \tau).
\end{equation}
We immediately have
\begin{equation}
  \label{eq:91}
  \vec q(t)\cdot\mathsf{k}\cdot\uvec q(t)
  = \dot\gamma q_x(t)q_y(t)/q(t)
\end{equation}
and choose $\vec q^{\perp}(t) = \vec q(t)\times[\vec
q(t)\cdot\mathsf{k}]$ perpendicular to $\vec q(t)$ and $\vec
q(t)\cdot\mathsf{k}$. Then
\begin{equation}
  \label{eq:92}
  \vec q(t)\cdot\mathsf{k}\cdot\uvec q^{\perp}(t)  = 0
\end{equation}
and using Eqs.~(\ref{eq:90}, \ref{eq:91}, \ref{eq:92}),
Eq.~(\ref{eq:88}) turns into
\begin{equation}
  \label{eq:93}
  \ddot\Phi_{\vec q}(t)  + \nu_{\vec q(t)}\dot\Phi_{\vec q}(t)
  + q^2(t)C^2_{\vec q(t)}\Phi_{\vec q}(t)
  + \int_0^td\tau m^{\parallel}_{\vec q}(t, \tau)\dot\Phi_{\vec q}(\tau)
  + 2\dot\gamma\frac{q_x(t)q_y(t)}{q^2(t)}\dot\Phi_{\vec q}(t) = 0,
\end{equation}
independent of $H_{\vec q}^{\perp}(t)$. Averaging Eq.~(\ref{eq:93})
over the directions of $\uvec q(t)$, the last term vanishes and we
arrive at Eq.~(\ref{eq:66}). Here we identify
$m_{\vec q}(t, \tau) \equiv m_{\vec q}^{\parallel}(t,\tau)/
q^2(t)C_{q(t)}^2$.

By the same line of arguments, we arrive at the equation of motion for
the transverse density-current correlator
\begin{equation}
  \label{eq:94}
  \frac{d}{dt}H_{\vec q}^{\perp}(t)
  + [\mu_{\vec q(t)} - \nu_{\vec q(t)}]H_{\vec q}^{\perp}(t)
  + \int_0^td\tau m_{\vec q}^{\perp}(t,\tau)H_{\vec q}^{\perp}(\tau) = 0.
\end{equation}
Note that it is implicitly coupled to $\Phi_{\vec q}(t)$ through the
memory kernel.

\section{Numerics}
\label{sec:numerics}

In order to solve Eqs.~(\ref{eq:64}, \ref{eq:66}, \ref{eq:67})
numerically, we adapted an established code \cite{fuchs+cates02}. The
temperature independence of the hard sphere equations is exploited and
the explicit dependence on the coefficient of restitution has been
added. For the structure factor, we use the Percus-Yevick
\cite{percus+yevick58} explicit solution for hard spheres
\cite{ashcroft+lekner66} and for the value of the pair correlation
function at contact, $\chi$, we use the Woodcock equation of state
\textsc{WC1} \cite{bannerman+lue10}. The collision frequency is then
given as the elastic Enskog expression,
$\omega_c(\varphi, T) \simeq 24\varphi\chi d^{-1}\sqrt{T/\pi}$
\cite{hansen+mcdonald90}. For simplicity, instead of calculating
$\sigma$ directly, we determine the temperature independent quantity
\begin{equation}
  \label{eq:95}
  \check\eta(\Pe, \varphi, \varepsilon) := \frac{1 + \varepsilon}{2}
  \int_0^{\infty}\frac{dt^*}{\sqrt{1 + (\Pe t^*)^2/3}}
  \int_0^{\infty}\frac{dk^*{k^*}^4}{60\pi^2}\times
  \frac{S'_{\overline k^*(-t^*)}S'_{k^*}}{S_{k^*}^2}
  \Phi^2_{\overline k(-t^*)}(t^*).
\end{equation}
From this result, the quantities of interest can be obtained as
\begin{equation}
  \label{eq:96}
  \frac{\sigma}{nT} = \frac\pi6\check\eta\frac\Pe\varphi,\quad
  \frac{\eta}{\eta_0(T)} = \frac{2\pi}{15\varphi\chi}\check\eta,\quad
  Bd = \frac{\pi}{576}\times\frac{\check\eta}{(\varphi\chi)^2\Pe}.
\end{equation}
As $\eta_0 \propto\sqrt T$, we can derive $\eta/\eta_0(T_0) =
\sqrt{T/T_0}\times\eta/\eta_0(T)$ and in the same manner
$\dot\gamma/\omega_c(T_0) = \Pe\sqrt{T/T_0}$.

The Enskog term, $\nu_q$ Eq.~(\ref{eq:56}), is known to drastically
underestimate sound damping at high densities. To this end we replace it,
\begin{equation}
  \label{eq:97}
  \nu_q \to \frac{20}{3}\times\frac{D_S}{d^2}[1 + 3j''_0(qd)],
\end{equation}
where we chose the Enskog expression for the sound damping constant $D_S$
\cite{garzo+montanero02} and note that this expression has the correct
hydrodynamic limit, $2D_Sq^2$ for $q\to0$.

For the wave number integrals, the wave numbers are discretized uniformly at
100 points between $k^* = 0.4$ and $k^* = 40$. The initial time step is
$\Delta\tau = 10^{-5}$ and the step size is doubled every 100 time steps to
bridge the time scales.

\section{Jenkins' Constitutive Equations}
\label{sec:jenk-const-equat}

\Citet{jenkins07} has developed a semi-empirical constitutive equation
for dense granular shear flow in the Bagnold regime and applicable for
densities $0.49\lesssim\varphi\lesssim0.6$. In our notation
it consists of an expression for the dimensionless shear stress
\begin{equation}
  \label{eq:98}
  \hat\sigma_{J1} := \frac{8J}{5\sqrt\pi}\left(
    \frac{30}{J}\times\frac{1-\varepsilon^2}{c_1}
  \right)^{1/3}G_1^{5/6}
\end{equation}
together with a dimensionless cooling rate
$\Gamma_{J1} := \frac{1 - \varepsilon^2}{L_1/d}$, modified from the Enskog
approximation by an effective dimensionless length scale
\begin{equation}
  \label{eq:99}
  L_1/d := \frac12\left[\frac{30}{J}(1-\varepsilon^2)c_1G_1\right]^{1/3}.
\end{equation}
Here $J := 1 + \pi/12$, $G_1\equiv\varphi\chi_{J1} := 0.85\varphi/(0.64 -
\varphi)$ is essentially an empirical pair correlation function made
to diverge at a presumed jamming density of $0.64$, and $c_1 \approx 0.75$ is
a fit parameter.

Using Eq.~(\ref{eq:96}), we can relate the Bagnold coefficient
\begin{equation}
  \label{eq:100}
  B_{J1}d = \frac{\varphi\hat\sigma_{J1}}{96G_1^2(\Pe_{\infty}^{J1})^2}
\end{equation}
to the dimensionless shear stress $\hat\sigma_{J1}$. Here the P\'eclet
number is determined consistently by Eq.~(\ref{eq:7}), $\Pe_{\infty}^{J1}
= \Gamma_{J1}/\hat\sigma_{J1}$, such that
\begin{align}
  \label{eq:101}
  B_{J1}d &= \frac{\varphi{\hat\sigma}_{J1}^3}{96G_1^2\Gamma_{J1}^2}
              = \frac{120\times30^{2/3}J^{4/3}}{375\pi^{3/2}c^{1/3}}
              (1 - \varepsilon^2)^{-1/3}\varphi G_1^{7/6}\\
  \label{eq:102}
            &\approx 0.833(1 - \varepsilon^2)^{-1/3}\varphi[\varphi\chi_{J1}(\varphi)]^{7/6}.
\end{align}

Later \citet{jenkins+berzi10} published a refined version of this
model where now
\begin{equation}
  \label{eq:103}
  \hat\sigma_{J2} := \frac{4(1+\varepsilon)}{5\sqrt\pi}\left(
    \frac{30}{c_2}\times\frac{1-\varepsilon^2}{1 + \varepsilon}
  \right)^{1/3}G_2^{8/9}
\end{equation}
and we have made the approximation
$J(\varepsilon) \simeq (1 + \varepsilon)/2$. Here
$G_2 \equiv \varphi\chi_{J2}(\varphi) := 0.63\varphi/(0.6 - \varphi)$
is another empirical pair correlation function and $c_2\approx 0.5$ is
a fit parameter. For the effective length scale they now propose
\begin{equation}
  \label{eq:104}
  L_2/d := \left(
    30\frac{1-\varepsilon^2}{1 + \varepsilon}c_2^2
  \right)^{1/3}G_2^{2/9}.
\end{equation}
Effectively we obtain
\begin{align}
  \label{eq:105}
    B_{J2}d &= \frac{\varphi{\hat\sigma}_{J2}^3}{96G_2^2\Gamma_{J2}^2}
    = \frac{4\times30^{2/3}c_2^{1/3}}{25\pi^{3/2}}
    \times\frac{(1 + \varepsilon)^{4/3}}{(1 - \varepsilon^2)^{1/3}}\varphi
                    G_2^{10/9}\\
  \label{eq:106}
                  &\approx 0.555\left(\frac{1+\varepsilon}{2}\right)^{4/3}
                    (1 - \varepsilon^2)^{-1/3}\varphi[\varphi\chi_{J2}(\varphi)]^{10/9},
\end{align}
where the main difference to Eq.~(\ref{eq:102}) is the different pair
correlation function $\chi_{J2}$ and the more complicated dependence
on the coefficient of restitution $\varepsilon$.

%

\end{document}